\documentclass[
    10pt
    ,aps
    ,pra
    ,oneside
    ,nobibnotes
    ,floatfix
    ,twocolumn
    ,nofootinbib
    ,superscriptaddress
    ,showkeys
]{revtex4-2}

\usepackage{amsmath,amssymb,bm}
\usepackage{array}
\allowdisplaybreaks[3]
\usepackage{siunitx}
\usepackage{graphicx}\graphicspath{{}{Graph2/}{Graph2/A0}{Graph2/A1}{Graph2/A2}
}

\usepackage[svgnames]{xcolor}

\usepackage{iftex}
\iftutex
    \usepackage{fontspec}

    \usepackage[math-style=ISO,bold-style=ISO]{unicode-math}
\else

    \usepackage[T1]{fontenc} 
    \usepackage[utf8]{inputenc} 
\fi

\usepackage[english, showlanguages]{babel}

    \iftutex
        
    \else
        
    \fi

    \DeclareMathOperator{\const}{const}

    \newcommand{\parder}[2]{\frac{\partial #1}{\partial #2}}

    \newcommand{\dif}[2][]{\mathop{}\!\mathrm{d}
        \if
            \relax\detokenize{#1}\relax
        \else
            ^{\mkern-1.mu#1}\mkern-2.5mu 
    \fi
    #2\,}
    \newcommand{\der}[2]{\frac{\dif{#1}}{\dif{#2}}}
    \newcommand{\tder}[2]{{\dif{#1}}/{\dif{#2}}}

    \newcommand{\mean}[1]{\left\langle #1\right\rangle}

\usepackage[
    colorlinks
        ,linkcolor=violet
        ,filecolor=purple
        ,citecolor=teal
        ,urlcolor=magenta
    ,unicode
    ,pdfpagemode=UseOutlines
    ,pdfdisplaydoctitle=true
    ,pdfpagetransition=Wipe
    ,pdfusetitle
]{hyperref}
\hypersetup{
        pdfkeywords={plasma, MHD stability, ballooning modes, mirror trap, Gas-Dynamic Trap}
}
\usepackage{bookmark}
\usepackage[justification=centerlast]{caption}

\begin{document}

\def\bot{\mathrel\perp}

\title{
    Wall stabilization of high-beta anisotropic plasmas in an axisymmetric mirror trap
}
\author{Igor Kotelnikov}
    \email{I.A.Kotelnikov@inp.nsk.su}
    \affiliation{Budker Institute of Nuclear Physics, Novosibirsk, Russia}
\author{Vadim Prikhodko}
    \email{V.V.Prikhodko@inp.nsk.su}
\author{Dmitri Yakovlev}
    \email{D.V.Yakovlev@inp.nsk.su}
    \affiliation{Budker Institute of Nuclear Physics SB RAS, Novosibirsk, 630090, Russia}
\date{\today}

\begin{abstract}
    The stabilization of ``rigid'' flute and ballooning modes $m = 1$ in an axisymmetric mirror trap with the help of an ideally conducting lateral both in the presence and in the absence of end MHD anchors is studied. The calculations were performed for an anisotropic plasma in a model that simulates the pressure distribution during the injection of beams of fast neutral atoms into the magnetic field minimum at a right angle to the trap axis. It was assumed that the lateral wall has the shape of a cylinder with a variable radius, so that on an enlarged scale it repeats the shape of the plasma column.

    It has been found that for the effective stabilization of the listed modes by an ideally conducting lateral wall, the parameter beta ($\beta$, the ratio of the plasma pressure to the magnetic field pressure) must exceed some critical value $\beta_{\text{crit}}$. When combined with a conducting lateral wall and conducting end plates imitating MHD end stabilizers, there are two critical beta values and two stability zones $0<\beta<\beta_{\text{ crit}1}$ and $\beta_{\text {crit}2}<\beta<1$ that can merge, making the entire range of allowable beta values $0<\beta<1$ stable.

    The dependence of the critical betas on the degree of plasma anisotropy, the mirror ratio, and the width of the vacuum gap between the plasma and the lateral wall is studied. In contrast to the works of other authors devoted to the plasma model with a sharp boundary, we calculated the boundaries of the stability zone for a number of diffuse radial pressure profiles and several axial magnetic field profiles.

\end{abstract}
\maketitle

\section{Introduction}\label{s1}

%
Continuing the study of ballooning instabilities in isotropic plasma started in our article \cite{Kotelnikov+2022NF_62_096025}, in this paper we present the results of the study of the so-called \emph{wall stabilization} of rigid flute and ballooning perturbations with azimuthal mode number $m=1$ in a mirror trap (also called open trap or linear trap) with anisotropic plasma. According to the physics of the case, wall stabilization of plasma with a sufficiently high pressure is achieved by exciting image currents in the conducting lateral walls of the vacuum chamber surrounding the plasma column. These currents are directed oppositely by the diamagnetic current in the plasma, and the opposite currents, as is known, are pushed apart. Such pushing returns the pop-up ``tongues'' of plasma back to the axis of the trap.

In order to avoid a possible misunderstanding of the significance of the stabilization of some mode $m=1$, we should probably recall what constitutes a ballooning instability in an open axially symmetric trap. It is generally accepted to think of ballooning perturbations as small-scale deformations of the plasma equilibrium with a large azimuthal mode number $m\gg1$. It has long been known that the finite Larmor radius (FLR) effects can stabilize small-scale flute-type perturbations if $m>1$ \cite{Rosenbluth+1962NFSuppl_1_143}. Ballooning perturbations are, in fact, related to flute perturbations; therefore, it is commonly believed that they are also stabilized by the FLR effects.

In its pure form, flute perturbations can be detected in a low-pressure plasma (that is, in the limit of $\beta \to 0$, where $\beta_{}$ is the ratio of plasma pressure to the magnetic pressure field). They represent a magnetic tube of field lines, which, together with the plasma captured in it, floats to the periphery of the plasma column without changing its shape and without distorting the magnetic field. The classical method for stabilizing flute disturbances is to attach end MHD ajchors (for example, of the cusp type) to the central section of the open trap, which ``clamp'' the ends of the force tube. The same ``clamping'' effect is achieved by placing conductive plates at the ends of the plasma column. If the ends of the magnetic tube are ``frozen'' into the end plates, the tube cannot float without bending and deforming the magnetic field. Thus, the flute perturbation is transformed into a balloon type perturbation. Deformation of the magnetic field requires energy, which must be withdrawn from the thermal energy of the plasma. Therefore ballooning instability is possible if plasma beta exceeds a certain threshold value; it is important to emphasize that this threshold value in axially symmetric open traps is very high, on the order of 60-70\% \cite{BushkovaMirnov1986VANT_2_19e, RyutovStupakov1981IAEA_1_119, Kotelnikov+2021PST_24_015102}.


The FLR effects are not capable of stabilizing flute and ballooning perturbations with $m=1$. However, they impose rigidity on such perturbations, so that the plasma density distribution does not change in each cross section of the plasma column. The plasma column bends, shifting from the trap axis to different distances in different sections. Stabilization of such a ``rigid'' mode $m=1$ would mean stabilization of all flute and ballooning perturbations, provided that the modes $m>1$ are stabilized by the FLR effects.

Unlike many other works on the stability of a rigid ballooning mode $m=1$
\cite{
    KaiserNevinsPearlstein1983PF_26_351, 
        Berk+1985PPCNFR_2_321,
        Berk+1984PF_27_2705, 
        HaasWesson1967PF_10_2245, 
        DIppolitoHafizi1981PF_24_2274, 
        DIppolitoMyra1984PF_27_2256,
    KaiserPearlstein1985PhysFluids_28_1003,
    Kesner1985NF_25_275,
    LiKesnerLane1985NF_25_907, 
            LiKesnerLane1987NF_27_101,
        LiKesnerLane1985NF_25_907, 
        LiKesnerLoDestro1987NF_27_1259
},
%
which were focused exclusively on the plasma model with a stepwise pressure profile along the radius, in our article \cite{Kotelnikov+2022NF_62_096025} we studied the stability of the $m=1$ mode in a plasma with a diffuse pressure profile, but then limited ourselves to the case of an isotropic plasma. Specifically, we calculated the so-called critical beta $\beta_{\text{crit}}$, such that the rigid flute and ballooning modes $m=1$ can be stabilized by the lateral conducting wall without using any other MHD stabilization methods if $ \beta > \beta_{\text{crit}}$. A second case was also analyzed when the lateral conductive wall was supplemented with conducting end plates installed in magnetic mirror throats, which should have simulated the installation of end MHD stabilizers, such as cusp. In this case, two stability zones were found: the first at small beta, $\beta < \beta_{\text{crit}1}$, the second at large beta, $\beta > \beta_{\text{crit}2}$ , and these two zones can merge.


%
In this article, we study wall stabilization on one particular model of anisotropic plasma, which allows us to perform a significant part of the calculations in an analytical form. The key equation for solving the problem was derived by Lynda LoDestro \cite{LoDestro1986PF_29_2329} in 1986, but, in fact, neither she nor anyone else has ever used this equation. We found few papers \cite{Devoto+1987NF_27_425, Dobrott+1987PF_30_2149, Molvik+1990NF_30_1061, ArseninKuyanov1996PPR_22_638} that refer to LoDestro's paper but do not use her equation. Probably oblivion for many years of LoDestro's work is due to the early termination of the TMX (Tandem Mirror Experiment) and MFTF-B (Mirror Fusion Test Facility B) projects in the USA in the same 1986 \cite{Ongena+2016NaturePhysics_12_398}. However, the achievement of high electron temperature and high beta in the gas dynamic trap (GDT) at the Budker Institute of Nuclear Physics in Novosibirsk
\cite{
    Ivanov+2003PhysRevLett_90_105002, 
    Simonen+2010JFE_29_558, 
    Bagryansky+2011FST_59_31,
    Bagryansky+2015PhysRevLett_114_205001,
    Bagryansky+2015NF_55_053009,
    Bagryansky+2016AIPConfProc_030015,
    Akhmetov+2018PPT_5_125,
    Yakovlev+2018NF_58_094001,
    Bagryansky+2019JFE_38_162,
    Shalashov+2022NF_62_124001},
emergence of new ideas \cite{Beklemishev2016PoP_23_082506} and new projects \cite{Granetzny+2018APS_CP11_150, Bagryansky+2020NuclFusion_60_036005, WHAM2020, Egedal+2022NF_62_126053, Yakovlev+2022NF_62_076017} makes us rethink old results.


Another reason for the loss of interest in the LoDestro equation could be the insufficient power of computers of that era. It is possible to calculate the critical beta for a plasma with a sharp boundary without this equation, which was done by LoDestro's predecessors. As the plasma model becomes more complex, the demands on computer performance and resources increase rapidly. It takes less than an hour on a modern desktop PC with eight cores to recalculate all the graphs presented in our article \cite{Kotelnikov+2022NF_62_096025} for the isotropic plasma model. But calculations for some anisotropic plasma models, which we have not yet published, take weeks and even months of continuous work.

%
To avoid unnecessary repetitions, we will not analyze the content of the articles cited above, since their detailed review was previously done in our article \cite{Kotelnikov+2022NF_62_096025}, and we will immediately proceed to the description of the new calculations.  In section \ref{s2}, LoDestro's equation will be written and the necessary notation will be introduced. In section \ref{s3}, the anisotropic pressure model used below will be formulated. The results of calculating some coefficients in the LoDestro equation for this model are presented in Appendices \ref{A1} and \ref{A2}. Section \ref{s4} presents the results of calculating the critical beta in the limit when the conducting wall surrounding the plasma column is almost conforms to the lateral boundary of the plasma column, but does not touch it. In this limit, LoDestro's ordinary differential equation of second order reduces to an integral along the $z$ coordinate on the trap axis; the integral turns to zero at the critical beta value. The section \ref{s5} describes solution of the LoDestro equation by the shooting method and presents the results of calculations for several model profiles of pressure and magnetic field. In section \ref{s6}, the shooting method is again used to solve the LoDestro equation with other boundary conditions that imitate the effect of conducting end plates or MHD anchors installed in magnetic plugs. The final section \ref{s9} summarizes our results and conclusions.

\section{LoDestro equation}\label{s2}

The LoDestro equation is a second-order ordinary differential equation for the function
    \begin{equation}
    \label{2:02}
    \phi(z) = a(z) B_{v}(z) \xi_{n}(z)
    ,
    \end{equation}
which depends on one coordinate $z$ along the trap axis and is expressed in terms of the variable radius of the plasma column boundary $a=a(z)$, the vacuum magnetic field $B_{v}=B_{v}(z)$ and the virtual small displacement $ \xi_{n}=\xi_{n}(z)$ of the plasma column from the axis. It is obtained on the assumption that
%
\begin{itemize}
   \item FLR effects are strong enough, $m=1$;
   \item the paraxial (long-thin) approximation applies;
   \item the pressure tensor depends on only two functions: $p_{\bot}(\psi,B)$, $p_{\|}(\psi,B)$;
   \item plasma beta is not necessarily small.
\end{itemize}
The LoDestro equation does not take into account the resistance of plasma and conductive walls. It is also possible that the assumption about the dominance of the FLR effects is violated near the boundary of the plasma column, which may result in an incorrect account of the plasma drifts and rotation.
%
%
Plasma rotation is taken into account by A.~Beklemishev's theory of vortex confinement \cite{Beklemishev+2010FST_57_351}, but it is limited by the zero beta limit. The resistance was taken into account by Kang, Lichtenberg, and Nevins \cite{KangLichtenbergNevins1987PF_30_1416} in an isotropic plasma model with a sharp boundary.

In its final form, the LoDestro equation reads
    \begin{multline}
    \label{2:01}
    0 = \der{}{z}
    \left[
        \Lambda + 1 - \frac{2\mean{\overline{p}}}{B_{v}^{2}}
    \right]
    \der{\phi}{z}
    \\
    +
    \phi
    \left[
        - \der{}{z}\left(
            \frac{B_{v}'}{B_{v}} + \frac{2a'}{a}
        \right)
        \left(
            1 - \frac{\mean{\overline{p}}}{B_{v}^{2}}
        \right)
    +
    \frac{\omega^{2}\mean{\rho}}{B_{v}^{2}}
    \right.
    \\
    \left.
    -
    \frac{2\mean{\overline{p}}}{B_{v}^{2}}\frac{a_{v}''}{a_{v}}
    -
    \frac{1}{2}\left(
            \frac{B_{v}'}{B_{v}} + \frac{2a'}{a}
    \right)^{2}
    \left(
        1 - \frac{\mean{\overline{p}}}{B_{v}^{2}}
    \right)
    \right]
    ,
    \end{multline}
where the derivative $\tder{}{z}$ in the first two lines acts on all factors to the right of it, and the prime ($'$) is a shortcut for $\tder{}{z}$.
%
Other notations are defined as follows
    \begin{gather}
    \label{2:03}
    \frac{a^{2}}{2} = \int_{0}^{\psi_{a}} \frac{\dif{\psi}}{B}
    ,\\
    \label{2:03a}
    \frac{r^{2}}{2} = \int_{0}^{\psi} \frac{\dif{\psi}}{B}
    ,\\
    \label{2:04}
        B^{2} = B_{v}^{2} -2p_{\bot}
    ,\\
    \label{2:07}
    a_{v}(z) = \sqrt{\frac{2\psi_{a}}{B_{v}(z)}}
    ,\\
    \label{2:05}
    \overline{p} = \frac{p_{\bot} + p_{\|}}{2}
    ,\\
    \label{2:06}
    \mean{\overline{p}}
    =
    \frac{2}{a^{2}}
    \int_{0}^{\psi_{a}} \frac{\dif{\psi}}{B}\,\overline{p}
    ,\\
    \label{2:09}
    \Lambda = \frac
    {
        r_{w}^{2} + a^{2}
    }{
        r_{w}^{2} - a^{2}
    }
    .
    \end{gather}
Parameter $\psi_{a}$, used here, has the meaning of the reduced (i.e.\ divided by $2\pi$) magnetic flux through the plasma cross section $\pi a^{2}$. It is related to the plasma radius $a=a(z)$ by Eq.~\eqref{2:03}.
Equation \eqref{2:03a} relates the radial coordinate $r$ and the magnetic flux $\psi$ through a ring of radius $r$ in the $z$ plane. The magnetic field $B=B(\psi,z)$, weakened by the plasma diamagnetism, in the paraxial (long-thin) approximation (i.e., with a small curvature of field lines) is related to the vacuum magnetic field $B_{v}=B_ {v }(z)$ by the transverse equilibrium equation \eqref{2:04}.

Function $\Lambda =\Lambda (z)$ is expressed in terms of the actual radius of the plasma/vacuum boundary $a=a(z)$ and the radius of the conducting cylinder $r_{w}=r_{w}(z)$, which surrounds the plasma column. In the remainder of the paper we assume that $\Lambda $ is a constant.
%
%
The case $\Lambda(z)=\const$ will be referred to below as a variant of the proportional chamber. The shape of the conducting walls of such a chamber on an enlarged scale repeats the shape of the plasma column, that is, $r_{w}/a-\const$. Such a chamber is difficult to manufacture, but the assumption $\Lambda(z)=\const$ greatly simplifies the solution of the LoDestro equation.
The larger the value of $\Lambda $, the closer the conducting lateral wall is to the plasma boundary. The $\Lambda \to \infty $ limit corresponds to the case when the conducting side wall is as close as possible to the plasma boundary, repeating its shape, but does not touch the plasma. The limit $\Lambda \to 1$ means that the lateral conducting wall is removed to infinity.

Kinetic theory predicts (see, for example, \cite{Newcomb1981JPP_26_529}) that the transverse and longitudinal plasma pressures can be considered as functions of $B$ and $\psi$, i.e.\ $p_{\bot}=p_{\bot}(B ,\psi)$, $p_{\|}=p_{\|}(B,\psi)$. In Eq.~\eqref{2:01}, one must assume that the magnetic field $B$ is already expressed in terms of $\psi$ and $z$, and therefore $p_{\bot}=p_{\bot}(\psi,z)$, $p_{\|}=p_{\|}(\psi,z)$. The angle brackets in Eq.~\eqref{2:01} denote the mean value of an arbitrary function of $\psi$ and $z$ over the plasma cross section. In particular, the average value $\mean{\rho}$ of the density $\rho=\rho(\psi,z)$ is calculated using a formula similar to \eqref{2:06}, and $\omega $ is the oscillation frequency.

The boundary conditions for Eq.~\eqref{2:01} and similar equations in the study of ballooning instability are traditionally set at the ends of the plasma column at the magnetic field maxima $B_{v}=B_{\max}$, where $B_{v}'=0$ and $p_{\bot}=p_{\|}=0$. In accordance with the geometry of actually existing open traps, it is usually assumed that the magnetic field is symmetrical with respect to the median plane $z=0$, and the magnetic mirrors (i.e., field maxima) are located at $z=\pm L$.

Traditionally, two types of boundary conditions are considered. In the presence of conducting end plates directly in magnetic mirrors, it is required that the boundary condition
    \begin{equation}
    \label{2:11}
    \phi = 0
    \end{equation}
%
be satisfied at $z=\pm L$. A similar boundary condition is usually used in studying the stability of small-scale ballooning disturbances, thereby modeling the presence of a stabilizing cell behind a magnetic mirror (see, for example, \cite{Kotelnikov+2021PST_24_015102}).

If the plasma ends are isolated, the boundary condition
    \begin{equation}
    \label{2:12}
    \phi' = 0
    \end{equation}
is applied.
As a rule, it implies that other methods of MHD stabilization in addition to stabilization by a conducting lateral wall are not used. It is this boundary condition \eqref{2:12} that was used earlier in the works on the stability of the $m=1$ ballooning mode.



The LoDestro equation \eqref{2:01} with boundary conditions \eqref{2:11} or \eqref{2:12} constitutes the standard Sturm-Liouville problem. At first glance, it may seem that the solution of such a problem is not difficult. However, the LoDestro equation has the peculiarity that its coefficients can be singular. In the anisotropic pressure model, which will be formulated in the next section, the singularity appears near the minimum of the magnetic field in the limit $\beta \to 1$. By some indications, it can be assumed that our predecessors were aware of the singularity problem. For example, in Kesner's article \cite{Kesner1985NF_25_275}, the graphs in Fig.~2 break off at $\beta\approx0.9$. Unfortunately, he did not leave us recipes for dealing with this singularity.

\section{Anisotropic pressure}\label{s3}

%
In publications on the stability of the rigid ballooning mode, two models of anisotropic pressure have previously been used. \emph{Kesner} in his paper \cite{Kesner1985NF_25_275}
%
indicates
that in the first model the transverse pressure in a nonuniform magnetic field $B\leq B_{\max}$ varies according to the law
    \begin{equation}
    \label{3:01}
    p_{\bot}\propto B_{\max}^{2}-B^{2},
    \end{equation}
while in the second model
    \begin{equation}
    \label{3:02}
    p_{\bot}\propto (B/B_{\max})^{2}(1-B/B_{\max})^{n-1}.
    \end{equation}
%
The first model approximately describes the pressure distribution in an open trap, which arises when beams of neutral atoms (NB) are injected into plasmas at a right angle to the trap axis into the minimum of the magnetic field. This is the so-called normal NB injection. The second model corresponds to oblique NB injection, which forms a population of so-called sloshing ions.

%
In the present paper, we will restrict ourselves to the first model for three reasons. Firstly, it roughly describes the current state of an experiment on the  Compact Axisymmetric Toroid (CAT) Budker INP \cite{Bagryansky+2016AIPConfProc_030015, Akhmetov+2018PPT_5_125, SudnikovSoldatkina2019AIPCP_2179_020026}. Secondly, in this model, the coefficients in the LoDestro equation can be calculated without using numerical integration. Thirdly, in the second model, the ultimate beta can actually be limited by either mirror or the firehose instabilities.

%
In our internal classification, the isotropic plasma variant is designated `A0', the anisotropic pressure \eqref{3:01}, which is formed during normal injection of atomic beams, is designated `A1'. The oblique injection simulated by the formula \eqref{3:02} with an arbitrary index $n$ is denoted by `A\emph{n}'. The two cases $n=2$ and $n=3$ will be dealt with in our next article and are abbreviated `A2' and `A3'.  Variations with conductive sidewall stabilization will be marked with the letters `LW' (from Lateral Wall), and variations with combined sidewall stabilization and conductive plates installed in the neck of the magnetic plug will be marked with the letters `CW' (from Combined Wall). The design of the lateral conductive wall in the form of a proportional chamber will be labeled `Pr' (for Proportional). Thus, the label `A1-LWPr' in the figure corresponds to the variant of plasma stabilization with anisotropic pressure of the `A1' type in a proportional conducting chamber without connecting end MHD stabilizers. These designations will be used in the following figures. The influence of the shape of the lateral wall on the stability of the rigid ballooning mode will be studied in a separate article. In particular, there will be a comparison of the stabilizing effect of a proportional with a straight conductive chamber, which is assigned the abbreviation `St' (Straight).

%
In kinetic theory \cite{Newcomb1981JPP_26_529} it is proved that if one of the two pressures $p_{\bot}$ and $p_{\|}$ is given as a function of $B$, then the other is uniquely determined using the parallel equilibrium equation. The latter can be rewritten in terms of the partial derivative with respect to $B$ for a constant magnetic flux $\psi$ as
    \begin{equation}
    \label{3:03}
    p_{\bot} = - B^{2}\parder{}{B}\frac{p_{\|}}{B}
    .
    \end{equation}
%
Another key result of the kinetic theory is the assertion that the function $p_{\bot}/B^{2}$ always decreases as $B$ increases, i.e.
    \begin{equation}
    \label{3:04}
    \parder{}{B}\frac{p_{\bot}}{B^{2}}
    \leq
    0
    .
    \end{equation}
 By an application of \eqref{3:03}, one can rewrite this last result as
    \begin{equation}
    \label{3:05}
    \parder{}{B}\frac{p_{\bot}+p_{\|}}{B}
    \leq
    0
    .
    \end{equation}
First, for stability against the so-called ``firehose mode'', it is required that
    \begin{equation}
    \label{3:06}
    p_{\|}-\frac{B^{2}}{2}
    \leq
    p_{\bot}+\frac{B^{2}}{2}
    .
    \end{equation}
Secondly, stability against the so-called  ``mirror mode'' implies that
    \begin{equation}
    \label{3:07}
    \parder{}{B}
    \left(
        p_{\bot}+\frac{B^{2}}{2}
    \right)
    >
    0.
    \end{equation}
%
All these equations are satisfied by the functions\footnote{
     The inequalities \eqref{3:04}, \eqref{3:05} and \eqref{3:06} are satisfied if $p_{0}>0$. The inequality \eqref{3:07} formally leads to the condition $2p_{0}<B_{s}^{2}$, while the transverse equilibrium equation \eqref{2:04} has a solution in the entire region $B_{v} <B_{s}$ under the obviously more stringent condition $2p_{0}<\min(B_{v}^{2})<B_{s}^{2}$.
}
    \begin{gather}
    \label{3:08}
    p_{\bot}(B,\psi)=
    p(\psi)
    \left(1-B^2/B_{s}^2\right)
    ,\\
    \label{3:09}
    p_{\|}(B,\psi)=
    p(\psi)
    \left(1-B/B_{s}\right)^{2}
    ,\\
    \label{3:10}
    \overline{p}(B,\psi) =
    p(\psi)
    \left(1-B/B_{s}\right)
    .
    \end{gather}
%
They describe the plasma pressure profile with a peak near the magnetic field minimum in the median plane of an open trap. Both functions $p_{\bot}$ and $p_{\|}$ simultaneously vanish at $B=B_{s}$. We assume that $B_{s}$ does not exceed the magnetic field $B_{\max}$ in magnetic mirrors, bearing in mind that there is some cold plasma in the region $B_{s} < B < B_{\max}$, but its pressure is negligible. Such a profile approximately describes the real pressure distribution in the CAT facility, where the field $B_{s}$ approximately corresponds to the stop point of sloshing ions.


%
In the case \eqref{3:08} Eq.~\eqref{2:04} can be solved and the magnetic field $B$ weakened by the diamagnetic effect can be explicitly expressed in terms of the vacuum field $B_{v}$:
    \begin{gather}
    \label{3:11}
    B(\psi,z)=
    B_{s}\sqrt{
        \frac{B_{v}^{2}(z)-2p(\psi)}{B_{s}^{2}-2p(\psi)}
    }
    .
    \end{gather}
%
In the second variant \eqref{3:02}, Eq.~\eqref{2:04} can be solved in analytical form only if $n=2$ or $n=3$, but in any case, numerical integration is necessary when calculating such functions as $a$ and $\mean{\overline{p}}$, while for the functions \eqref{3:06}–\eqref{3:08} these integrals can be calculated analytically.

The pressure functions expressed in terms of $B_{v}$ will be denoted by the capital letter $P$,
    \begin{gather}
    \label{3:14}
    P_{\bot}(B_{v},\psi)=
    p(\psi)
    \frac{B_{s}^2-B_{v}^2}{B_{s}^2-2p(\psi)}
    ,\\
    \label{3:15}
    P_{\|}(B_{v},\psi)=
    p(\psi)
    \left(
        1
        -
        \sqrt{\frac{B_{v}^2-2p(\psi)}{B_{s}^2-2p(\psi)}}
    \right)^{2}
    ,\\
    \label{3:16}
    P(B_{v},\psi)=
    p(\psi)
    \left(
        1
        -
        \sqrt{\frac{B_{v}^2-2p(\psi)}{B_{s}^2-2p(\psi)}}
    \right)
    .
    \end{gather}
%
Turning to dimensionless variables, we will take the magnetic field at the stopping point $B=B_{v}=B_{s}$ as a unit of measurement, and the vacuum mirror ratio $B_{s}/\min(B_{v})$ at the stopping point will be further denoted via $R$. Then, keeping the same notation for dimensionless quantities as for dimensional ones, we have
    \begin{equation}
    B_{v0}=\min(B_{v})=1/R,
    \qquad
    B_{s}=1.
    \end{equation}
%
The unit of length along the radius $r$ is further fixed by the fact that we take the magnetic flux $\psi_{a}$ through the plasma section as a unit of measurement, assuming further that $\psi_{a}=1$. As for the coordinate $z$ along the trap axis, it will be further normalized to the distance $L$ between the median plane $z=0$ and the magnetic mirror. With this normalization, it turns out that the plugs are located in the planes $z=\pm 1$.

Below we represent the dependence of pressure on the magnetic flux $\psi$ as
    \begin{equation}
    \label{3:21}
    p(\psi)=p_{0}f_{k}(\psi)
    ,
    \end{equation}
%
where the dimensionless function $f_{k}(\psi)$ is defined as
    \begin{equation}
    \label{3:22}
    f_{k}(\psi) =
    \begin{cases}
      1 - \psi^{k}, & \mbox{if } 0\leq \psi \leq 1 \\
      0, & \mbox{otherwise}
    \end{cases}
    \end{equation}
%
for integer values of index $k$, and for $k=\infty$
is expressed in terms of a $\theta$-function such that $\theta(x)=0$ for $x<0$ and $\theta(x)=1$ for $x>0$:
    \begin{equation}
    \label{3:23}
    f_{\infty}(\psi) = \theta(1 - \psi)
    .
    \end{equation}
%

Parameter beta $\beta $ is defined as the maximum of the ratio $2p_{\bot}/B_{v}^{2}$,
    \begin{gather}
    \beta=\max(2p_{\bot}/B_{v}^{2})
    .
    \end{gather}
The maximum is reached on the trap axis (where $\psi=0$) at the minimum of the vacuum field (where $B_{v}=1/R$), so that
in dimensionless notation
    \begin{gather}
    \label{3:18}
    p_{0}
    =
    \frac{\beta}{2 \left(\beta +R^2-1\right)}
    .
    \end{gather}
%
Parameter $p_{0}$ can vary within the range $0<p_{0}<1/2R^{2}$, and $p_{0}\to 1/2R^{2}$ at $\beta\to 1$. At $\beta >1$, plasma equilibrium is impossible, since Eq.~\eqref{2:04} does not have a continuous solution.

As in our first article \cite{Kotelnikov+2022NF_62_096025}, the subsequent calculations were performed in the Wolfram \emph{Mathematica}$^{\copyright}$ for four values of the index $k=\{1,2,4,\infty \}.$
The $f_{1}$ function describes the smoothest pressure profile. For $\beta_{}/B_{v}^{2}\ll1$ it approximately gives the parabolic dependence of the pressure $p$ on the coordinate $r$. The larger the index $k$, the more table-like distribution $p$ near the axis of the plasma column and the steeper it is near the boundary of the column. Index $k=\infty $ corresponds to a pressure profile in the form of a step with a sharp boundary.

%
For the above radial pressure profiles, Wolfram \emph{Mathematica}$^{\copyright}$ was able to calculate integrals \eqref{2:03} and \eqref{2:06} analytically (with the help of some human prompts as described in Appendix \ref{A3}). The results of the calculations are summarized in Appendixes \ref{A1} and \ref{A2}.


An important characteristic of the plasma is the degree of its anisotropy. If the latter is formally defined as
     \begin{equation}
     \label{3:24}
     A
     =
     \frac{
         P_{\bot}(B_{v0},0)-P_{\|}(B_{v0},0)
     }{
         P_{\bot}(B_{v0},0)+P_{\|}(B_{v0},0)
     }
     ,
     \end{equation}
relating it to the plasma pressure at the minimum of the magnetic field on the trap axis, we obtain
     \begin{equation}
     \label{3:25}
     A = \frac{\sqrt{1-\beta }}{R}
     .
     \end{equation}
This shows that the anisotropy is the greater, the smaller $R$, reaching a maximum at $R\to 1$. In this connection, the parameter $R$ will sometimes be called the anisotropy parameter. The limit $R\to\infty$ corresponds to the case of an isotropic plasma. Calculating this limit in the formulas collected in Appendices \ref{A1} and \ref{A2}, one can re-derive the formulas for isotropic plasma published earlier in Ref.~\cite{Kotelnikov+2022NF_62_096025}. However, in this article we restrict ourselves to the case of $R\leq M$, since the plasma pressure should tend to zero at the neck of the magnetic mirror, where dimensionless $B_{v}=M/R$ and $M=\max(B_{v})/\min(B_{v})$ is the mirror ratio. We also note that the parameter $R$ characterizes the spatial width of the plasma pressure distribution along the magnetic field lines. The larger $R$, the wider the axial pressure profile.

\section{Thin vacuum gap}
\label{s4}

As the lateral conducting wall approaches the plasma/vacuum boundary, where it produces its maximum stabilizing effect, parameter $\Lambda $ tends to infinity. In the limit $\Lambda \to \infty$, it is possible to make analytic progress in solving Eq.~\eqref{2:01}, and, hence, in assessing the effects of a diffuse profile and anisotropy on the rigid ballooning mode stability.

%
Stabilization of the $m=1$  mode by a conducting wall in the $\Lambda \to \infty $ limit was previously studied by \emph{Kesner} \cite{Kesner1985NF_25_275}, \emph{Li, Kesner and Lane} \cite{LiKesnerLane1985NF_25_907} in the case of a plasma with a sharp-boundary radial profile.
In particular, apart of isotropic plasma \emph{Kesner} analyzed anisotropic pressure model \eqref{3:02} and concluded the critical beta becomes smaller as the degree of anisotropy increases.

%
As shown in Ref.~\cite{Kotelnikov+2022NF_62_096025}, in the $\Lambda \to \infty$ limit, the Sturm-Liouville problem \eqref{2:01}, \eqref{2:12} reduces to solving the integral equation
    \begin{multline}
    \label{4:02}
    \omega^{2}
    \int_{-1}^{+1}
    \frac{\mean{\rho}}{B_{v}^{2}}
    \dif{z}
    =
    \\
    =
    \int_{-1}^{+1}
    \left[
        \frac{2\mean{\overline{p}}}{B_{v}^{2}}\frac{a_{v}''}{a_{v}}
        +
        \frac{1}{2}\left(
            \frac{B_{v}'}{B_{v}} + \frac{2a'}{a}
        \right)^{2}
        \left(
            1 - \frac{\mean{\overline{p}}}{B_{v}^{2}}
        \right)
    \right]
    \dif{z}
    .
    \end{multline}
It allows one to calculate the squared oscillation frequency $\omega^{2}$ if the radial profile of pressure $\overline{p}=(p_{\bot}+p_{\|})/2$, density $\rho$, and vacuum magnetic field $B_{v}$ are given. At the margins of the stable regime, the oscillation frequency is equal to zero, $\omega^{2}=0$.
In the stability region $\omega^{2}>0$, and instability takes place if $\omega^{2}<0$.

%
In a plasma with anisotropic pressure, condition \eqref{3:06} together with Eq.~\eqref{2:04} guarantees that $\overline{p}<B_{v}^{2}/2$. Hence, the multiplier
$(1-{\overline{p}}/{B_{v}^{2}})>0$ and the second term in the square brackets are always positive. Therefore, the rigid ballooning mode in a plasma with anisotropic pressure can also be stabilized by a conducting wall located close enough to the lateral surface of the plasma column.

%
Further in this section, we present the results of calculating the critical value of beta in anisotropic plasma in the limit $\Lambda = \infty $. The calculation method is described in detail in our previous article using the example of isotropic plasma \cite{Kotelnikov+2022NF_62_096025}. Here, we restrict ourselves to indicating that the critical value $\beta_{\text{crit}}$ of the parameter $\beta_{}$, corresponding to the marginal stability $\omega^{2}=0$, is to be determined as a root of the equation
    \begin{equation}
    \label{4:10}
    W(\beta) = 0
    ,
    \end{equation}
where $W(\beta)$ denotes the integral on the right-hand-side of Eq.~\eqref{4:02}.
%
Some peculiarities of searching for roots of this strongly non-linear equation in the Wolfram \emph{Mathematica}$^{\copyright}$ are described in Appendix \ref{A3}.

As in Ref.~\cite{Kotelnikov+2022NF_62_096025}, we have performed calculations for two vacuum magnetic field models. However, below we present the results only for the second model, since the first model revealed approximately the same trends as the second one.

\begin{figure}
  \centering
  \includegraphics[width=\linewidth]{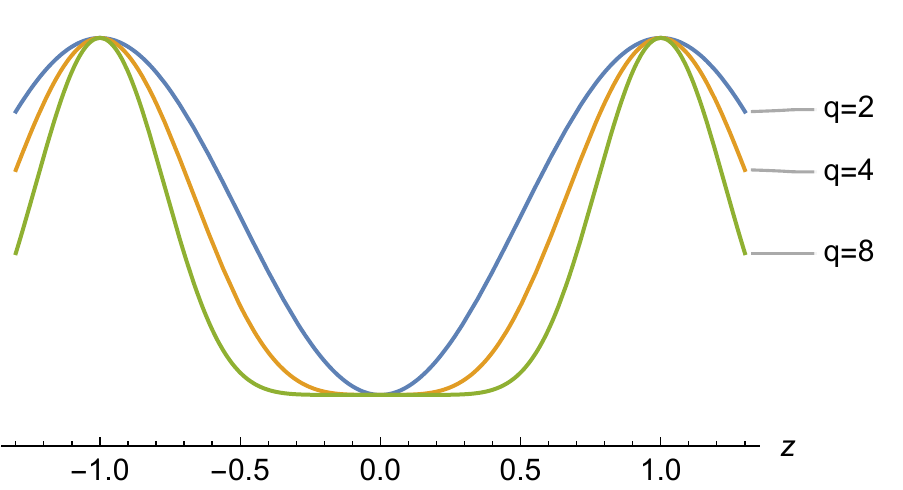}
  \caption{
    Axial profile of the vacuum magnetic field \eqref{4:35} with mirror ratio $M=8$ and different indices $q$ indicated on the graphs.
  }\label{fig:Bv_vs_z_q}
\end{figure}

In the second model, the vacuum magnetic field was given by a two-parameter family of functions
    \begin{equation}
    \label{4:35}
    B_{v}(z) =
    \left[
        1 + (M-1)\sin^{q}(\pi z/2)
    \right]/R
    \end{equation}
with five values of the mirror ratio $M=\{24,16,8,4,2\}$ and three values of the index $q=\{2,4,8\}$; the latter determines the width and slope of the magnetic mirrors, as shown in the figure~\ref{fig:Bv_vs_z_q}. The calculations were carried out for discrete values of the parameter $R$ in the range from $R=1.1$ to $R=M$. As noted above, the parameter $R$ can serve as a characteristic of plasma anisotropy. According to the equation~\eqref{3:25}, the closer $R$ is to unity, the greater the degree of anisotropy.

\begin{figure*}
  \centering
%
\includegraphics[width=\linewidth]{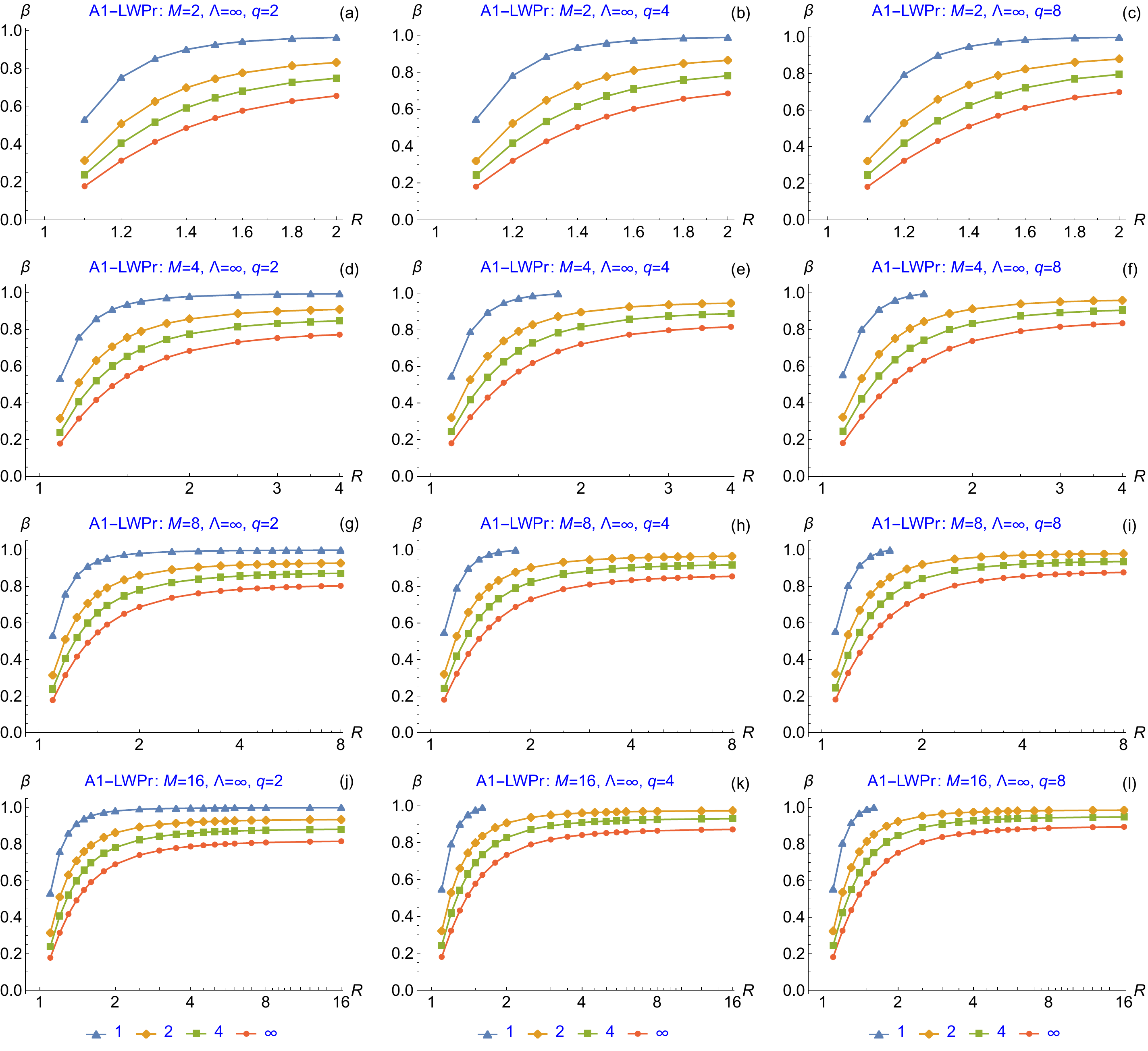}
  \caption{
    Critical beta in the model magnetic field \eqref{4:35} versus the anisotropy parameter $R$ for different mirror ratios $M$ and different indices $q$ in the  limit $\Lambda \to \infty $. The stability zone for the radial profile with index $k$ is located above the corresponding curve.
  }
  \label{fig:GMK-beta_vs_R-LInf}
\end{figure*}

Results of calculations are summarized in series of graphs in Fig.~\ref{fig:GMK-beta_vs_R-LInf}. Within each graph, it is not difficult to detect a trend towards a decrease in the critical value of beta with an increase in the steepness of the radial pressure profile as the index $k$ increases from $k=1$ to $k=\infty$ for a fixed pair of parameters $q$ and $R$.

Second trend is that critical beta rapidly approaches zero at $R\to1$, but in our calculations we did not take $R$ values less than $1.1$, believing that they are unlikely can be achieved experimentally. At $R=1.1$, critical beta ranged from $0.531$ for the smoothest radial profile $k=1$ to $0.178$ for a plasma with a sharp boundary at $k=\infty$. For sufficiently large values of $R\geq 3\div 5$, critical beta approaches unity, and the closer the smaller  parameter $k$. It is significant that with an increase in $R$, the stability zone disappears for smooth radial profiles (primarily for $k=1$).

The dependence on the index $q$, which characterizes the width and steepness of the magnetic mirrors (the larger $q$, the narrower the mirror width), does not seem to be very significant, except that profiles with an increased $q$ are more prone to the disappearance of the stability zone as the parameter $R$ increases. In other words, stabilization of the rigid ballooning mode is more problematic in traps with short and steep magnetic mirrors.

We also noticed a moderate dependence of the critical beta on the mirror ratio $M$. Decreasing $M$ from $16$ to $4$ reduces the value of $\beta_{\text{crit}}$ to the second or third decimal place.

\begin{figure*}
  \centering
  \includegraphics[width=0.3\linewidth]{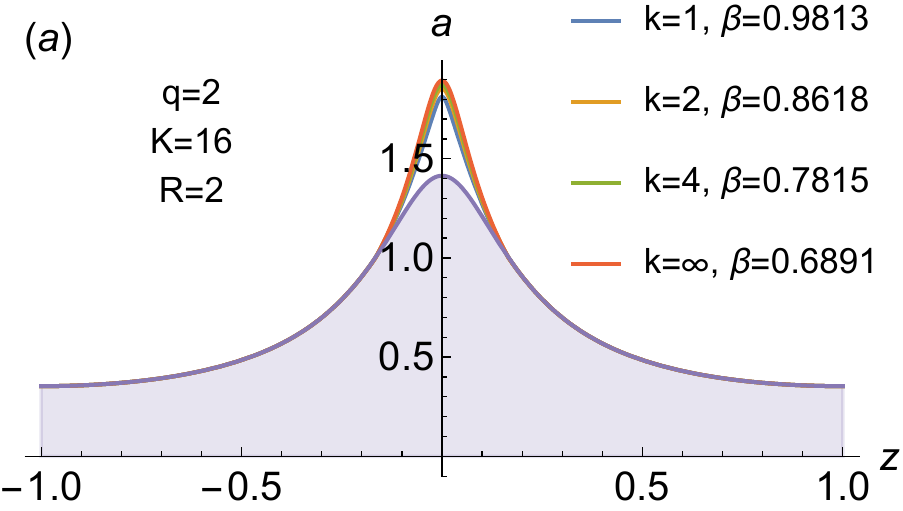}\hfil
  \includegraphics[width=0.3\linewidth]{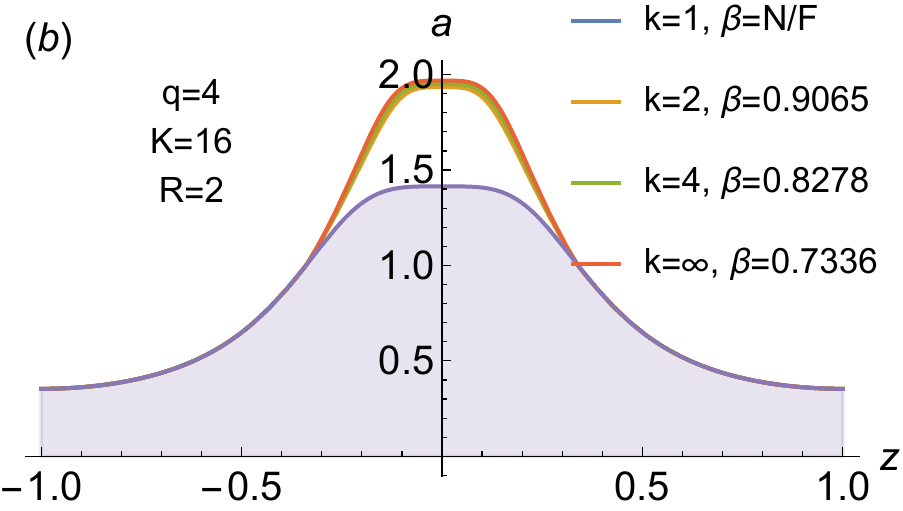}\hfil
  \includegraphics[width=0.3\linewidth]{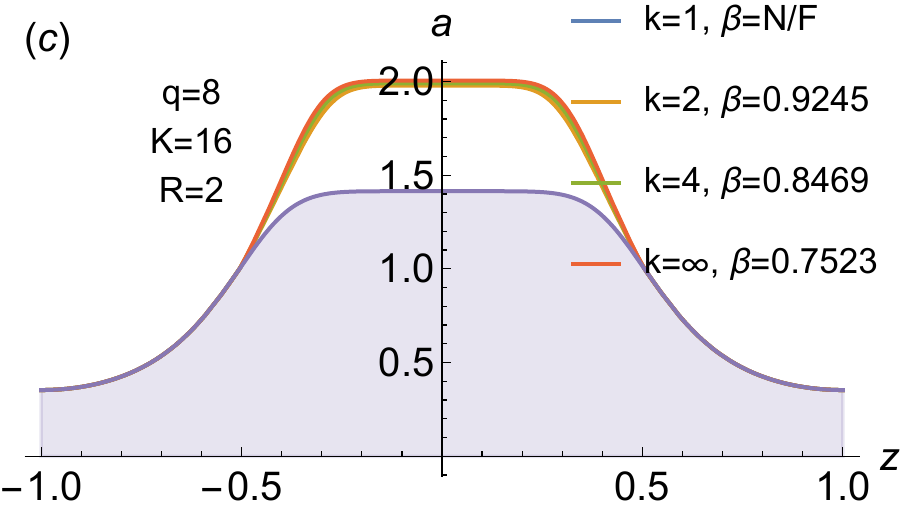}
  \caption{
    Axial profile of the plasma boundary in the model magnetic field \eqref{4:35} for  $\Lambda \to \infty $, $R=2$, $M=16$, different parameters $q$ (shown in the graphs) and critical values of beta for different pressure profiles $k$ (also shown in the graphs). The area occupied by plasma at $\beta=0$ is shaded.
  }\label{fig:GM-KA-a_vs_z}
    \bigskip
  \centering
  \includegraphics[width=0.3\linewidth]{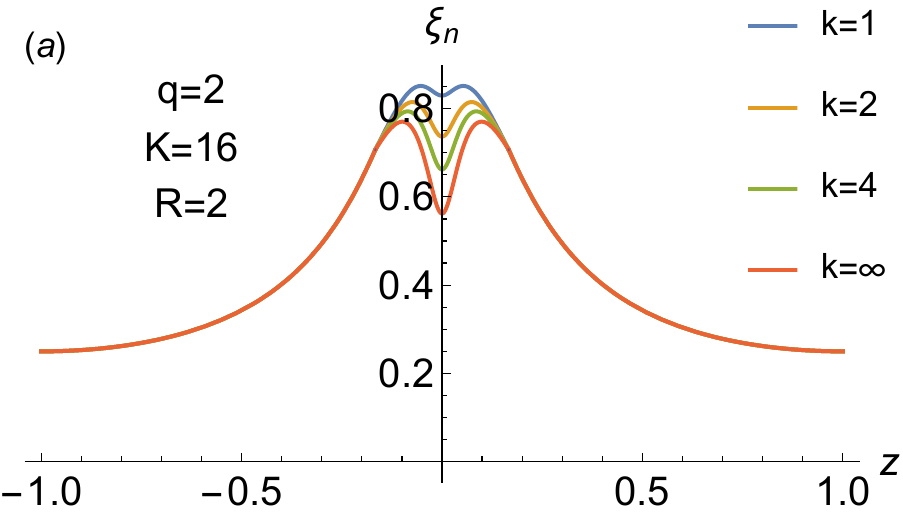}\hfil
  \includegraphics[width=0.3\linewidth]{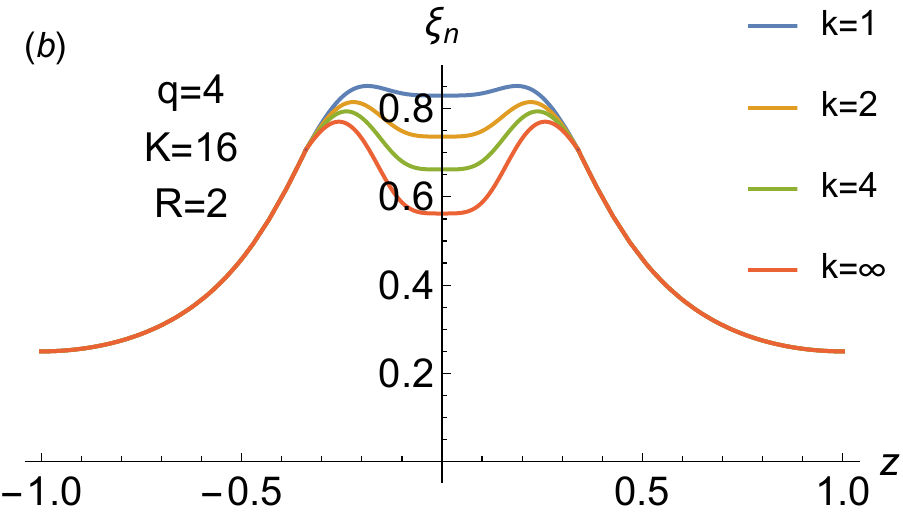}\hfil
  \includegraphics[width=0.3\linewidth]{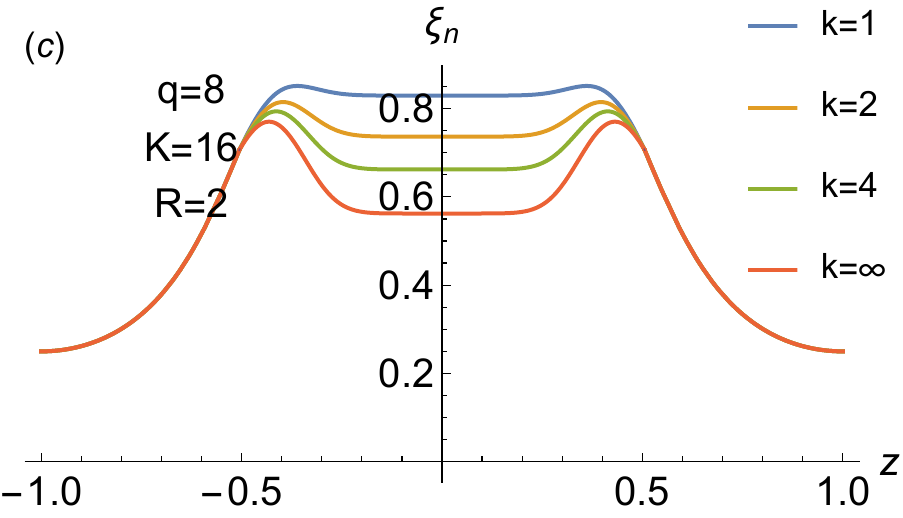}
  \caption{
    Axial profile of the displacement $\xi_{n}(z)$ of the plasma column in the model magnetic field \eqref{4:35} for $\beta =0.9$, $\Lambda = \infty $, $R=2$, $M=16$ and various values of $q$ and $k $ (indicated in the graphs).
  }
  \label{fig:GM-KA-ksi_vs_z}
\end{figure*}

Although function \eqref{4:35} is everywhere smooth, on the profile of the plasma boundary $a(z)$ near the median plane $z=0$, a ``swell'' is formed in the form of a ``thorn'' with a large curvature on point. An example of such a ``spike'' for $q=2$ is shown in Fig.~\ref{fig:GM-KA-a_vs_z}(a). At $q=8$, the ``thorn'' expands, forming a diamagnetic ``bubble'' named after Beklemishev, as in Fig.~\ref{fig:GM-KA-a_vs_z}(c). The graphs of the plasma boundary $a(z)$ in Fig.~\ref{fig:GM-KA-a_vs_z} are plotted for different values of $k$ and different values of $\beta_{\text{crit}}$ corresponding to them, but with the same value of the magnetic flux $\psi=\psi_{a}=1$ captured into the plasma. Interestingly, such plots of $a(z)$ almost coincide, although the values of $\beta_{\text{crit}}$ for radial pressure profiles with different $k$ differ quite significantly. Similar effect was reported in Ref.~\cite{Kotelnikov+2022NF_62_096025} for isotropic plasma. Note in passing that this effect simplifies the production of a proportional chamber.

The formation of a region of large curvature at $\beta \to 1$ leads to a violation of paraxial approximation, which was assumed in the derivation of the LoDestro equation. The formal condition for the applicability of paraxial approximation can be written as
    \begin{equation}
    \label{4:41}
    \psi_{a}\der{^{2}a^{2}_{k}}{z^{2}} \ll 1
    ,
    \end{equation}
where the functions $a^{2}_{k}$ for the pressure profiles with different indices $k$ are defined by Eqs.~\eqref{A1:1} in Appendix \ref{A1}. According to their definition \eqref{2:03} functions $a^{2}_{k}$ are proportional to the magnetic flux $\psi_{a}$ trapped in the plasma, but when writing Eqs.~\eqref{A1:1}  parameter $ \psi_{a}$ was set to one, so it was added explicitly to the condition \eqref{4:41}. Formulating the paraxiality condition, we assume that $\psi_{a}=a_{v0}^{2}B_{v0}/2$, where $a_{v0}$ is the radius of the plasma column in the median plane of a vacuum magnetic field at $\beta =0$.

An analysis of the graphs of the left-hand-side of the condition \eqref{4:41} for the magnetic field  \eqref{4:35} shows that it is most difficult to fulfill this condition in the case of $q=2$, when the maximum curvature is reached in the median plane $z=0 $. In the other two variants $q=4$ and $q=8$, the first two derivatives of the vacuum field $B_{v}(z)$ are equal to zero at $z=0$, so the curvature peak is formed at some distance from the median plane, where the vacuum field is larger and the local beta value is smaller. For the field of the type $q=2$ near the median plane, we approximately have
    \begin{equation}
    \label{4:42}
    B_{v}=B_{v0}\left[
        1+z^{2}/2l^{2}
    \right],
    \end{equation}
%
where $l$ determines the scale of the change in the vacuum field in this region. Substituting \eqref{4:42} into Eqs.~\eqref{A1:1} and calculating the right-hand-side of Eq.~\eqref{4:41} at the point $z=0$, we obtain the desired constraint on the paraxiality parameter. For the smoothest pressure profile $k=1$, this method yields
    \begin{subequations}
    \label{4:43}
    \begin{equation}
    \label{4:43-1}
    \frac{2 \sqrt{R^2-1}}{\sqrt{1-\beta } R}
    \frac{a_{v0}^{2}}{l^{2}}
    \ll 1
    .
    \end{equation}
For the other three cases ($k=\{2,4,\infty \}$), respectively, we have
    \begin{gather}
    \label{4:43-2}
    \frac{\sqrt{R^2-1}}{(1-\beta )R}
    \frac{a_{v0}^{2}}{l^{2}}
    \ll 1
    ,\\
    \label{4:43-4}
    \frac{
        2\sqrt{R^2-1} \Gamma\left({5}/{4}\right)^2
    }{
        \sqrt{\pi } (1-\beta)^{5/4} R
    }
    \frac{a_{v0}^{2}}{l^{2}}
    \ll 1
    ,\\
    \label{4:43-inf}
    \frac{\sqrt{R^2-\beta }}{(1-\beta )^{3/2} R}
    \frac{a_{v0}^{2}}{l^{2}}
    \ll 1
    .
    \end{gather}
    \end{subequations}
%
This shows that the most severe restriction on the value of beta takes place for the steepest pressure profile ($k=\infty $). From Eq.~\eqref{4:43-inf} we obtain the condition
    \begin{equation}
    1-\beta \gg \left(
        {a_{v0}}/{l}
    \right)^{4/3}
    .
    \end{equation}
%
From the formal point of view of a refined mathematician, it can be true even for $\beta \to 1$ if the parameter $a_{v0}/l$ tends to zero even faster, but in real open traps the parameter $a_{v0}/l $, although small, has a finite value. Accordingly, our results will be unreliable if $\beta $ is too close to one.

%
The displacement profiles of the plasma column $\xi_{n}(z)$ are shown in Fig.~\ref{fig:GM-KA-ksi_vs_z}  for all radial pressure profiles $k$ at the same value $\beta=0.9$. We emphasize that the displacement is not constant, although, as shown in \cite{Kotelnikov+2022NF_62_096025}, $\phi(z)=\const$ for $\Lambda \to\infty$. At the critical values of beta indicated in Fig.~\ref{fig:GM-KA-a_vs_z}, the displacement profiles would be practically the same for all $k$, since the profiles of the plasma boundary $a(z) $ in Fig.~\ref{fig:GM-KA-a_vs_z} practically coincide.

Next section \ref{s5} describes the results of solving the Sturm-Liouville problem for the LoDestro equation \eqref{2:01} with a finite value of $\Lambda $. We used the $\Lambda \to\infty$ limit to test convergence of this solution.

\section{Finite vacuum gap}\label{s5}

We used the \texttt{Parametric\-NDSolve\-Value} built-in utility to find solution to the ordinary differential equation \eqref{2:01} in the Wolfram \emph{Mathematica}$^{\copyright}$ system. The utility returns a reference $pf$ to an interpolation function of the $z$ coordinate, which also depends on the free parameters $\beta_{}$, $\Lambda$, and $\omega $. Other parameters ($M$, $R$, $q$) were given by numbers.

Taking into account the symmetry of the magnetic field with respect to the median plane $z=0$, it suffices to find a solution to the equation at half the distance between the magnetic mirrors, for example, on the interval $0<z<1$. Due to the symmetry, the desired function $\phi(z)$ must be even, hence
    \begin{equation}
    \label{5:01}
    \phi'(0)=0.
    \end{equation}
%
As for the point $z=1$, then, according to \eqref{2:12}, there should be satisfied the equality
    \begin{equation}
    \label{5:02}
    \phi'(1)=0,
    \end{equation}
%
which corresponds to the case when the plasma ends are isolated from the conducting elements of the structure of the vacuum chamber. An obvious fit to the LoDestro equation with boundary conditions \eqref{5:01} and \eqref{5:02}  is the trivial solution $\phi\equiv0$. To eliminate the trivial solution, we impose the normalization condition
    \begin{equation}
    \label{5:03}
    \phi(0)=1.
    \end{equation}
%
It is possible to simultaneously satisfy the three boundary conditions \eqref{5:01}–\eqref{5:03} when solving a second-order ordinary differential equation only for a certain combination of the parameters $\Lambda $, $\beta $ and $\omega $. If a pair of parameters $\Lambda$ (geometry of the conducting wall) and $\beta$ (plasma pressure) is given, the solution of the Sturm-Liouville  problem \eqref{2:01}, \eqref{5:01}–\eqref{5:03} gives eigenfrequency  $\omega $ of the rigid ballooning mode. We will be interested in solving this problem for the case of proportional chamber when the ratio $r_{w}/a$ is constant (so that the function $\Lambda(z) $ is also a constant independent of $z$), and the oscillation frequency is equal to zero, $\omega=0$. This solution gives critical value of beta, $\beta_{\text{crit}}$, as a function of $\Lambda $. The application of the shooting method for solving such a problem in the Wolfram \emph{Mathematica}$^{\copyright}$ system was described earlier in Ref.~\cite{Kotelnikov+2022NF_62_096025} using the example of an isotropic plasma. Let us dwell on the features of the calculation for an anisotropic plasma.

%
In what follows, we denote by $z_{s}$ the coordinate of the stop point on the $z$ axis, where $B=B_{v}=B_{s}=1$, and $p_{\bot}=p_{\|}=0$.
%
For the second magnetic field model
    \begin{equation}
    \label{5:12}
    z_{s}
    =
    \frac{2}{\pi }\,
    \arcsin\left(
        \left(
            \frac{R-1}{M-1}
        \right)^{{1}/{q}}
    \right)
    .
    \end{equation}
%
In our model of anisotropic plasma, its pressure is zero in the region $z_{s}<z<1$ between the stop point $B_{v}=1$ and the magnetic mirror throat $B_{v}=M/R$.  At zero pressure, the LoDestro equation \eqref{2:01} takes an extremely simple form
    \begin{equation}
    \label{5:13}
    0 = \der{}{z}
    \left[
        \Lambda + 1
    \right]
    \der{\phi}{z}
    .
    \end{equation}
Its solution is the equality
    \begin{equation}
     \label{5:14}
    \left[
        \Lambda(z) + 1
    \right]
    \phi'(z)
    =\const,
    \end{equation}
where we allow for a moment that $\Lambda $ may be a function of  $z$, bearing in mind a future analysis of a conducting wall with different shape.
%
The constant on its right-hand side can be found from the boundary condition at $z=1$, where the derivative $\phi'(1)$ is equal to zero. It follows that the constant on the right-hand side of Eq.~\eqref{5:14} is also zero. Since the factor $\Lambda(z) + 1$ is greater than zero everywhere, we conclude that $\phi'(z)=0$ in the entire region $z_{s} < z < 1$. Thus, it suffices to find the numerical solution of the original equation \eqref{2:01} in the region $0 < z < z_{s}$.

It should be taken into account that the derivative $\phi'(z)$ undergoes a jump at $z=z_{s}$. Indeed, integrating Eq.~\eqref{2:01} over an infinitesimal neighborhood of the point $z_{s}$ from $z_{s}^{-}$ to $z_{s}^{+}$, we obtain the equation
    \begin{equation}
    \label{5:15}
    \left[ \Lambda(z_{s}) + 1 \right]
    \left[
        \phi'({z_{s}^{+}})
        -
        \phi'({z_{s}^{-}})
    \right]
    =
    \left[
        Q({z_{s}^{+}})
        -
        Q({z_{s}^{-}})
    \right]
    \phi(z_{s})
    ,
    \end{equation}
%
in which we took into account that $\Lambda $, $\phi$ and $\mean{\overline{p}}$ are continuous at the point $z=z_{s}$, in contrast to the derivative $\phi'(z)$ and the coefficient
    \begin{equation}
    \label{5:16}
    Q(z) =
    \frac{B_{v}'}{B_{v}}
    +
    \frac{2a'}{a}
    =
    \frac{2a'}{a}
    -
    \frac{2a_{v}'}{a_{v}}
    .
    \end{equation}
%
The jump in the coefficient is due to the fact that for $B=B_{v}=R$ the derivative of functions \eqref{3:08} and \eqref{3:10} jumps. Since $\phi'({z_{s}^{+}})=0$ and $Q({z_{s}^{+}})=0$, from Eq.~\eqref{5:15} we find the value that the derivative of $\phi'(z)$ must have at the point $z_{s}^{-}$ on the right boundary of the interval $0<z<z_{s}$ from its inner side:
    \begin{equation}
    \label{5:17}
    \phi'({z_{s}^{-}})
    =
    \frac{
        Q({z_{s}^{-}})
    }{
        \Lambda(z_{s}) + 1
    }\,
    \phi(z_{s})
    .
    \end{equation}
%
When solving Eq.~\eqref{2:01} on the interval $0<z<z_{s}^{-}$, the boundary condition \eqref{5:17} should be used instead of \eqref{5:02}.

Our calculations were done in the magnetic field \eqref{4:35} for the mirror ratios $M=\{4,8,16\}$, those combinations of parameters $k=\{1,2,4,\infty\}$, $q=\{2,4,8\}$, which are listed in section \ref{s4}, and for discrete constant values of $\Lambda=\{1,1.001,1.002,1.003, \ldots, 400,450,500\}$ in the interval from $\Lambda =1$ to $\Lambda=500$. For $\Lambda=500$, the critical beta value we calculated differed from the value found in previous section for $\Lambda=\infty$ only in the third decimal place, while in the isotropic plasma model the difference was observed only in the fifth decimal place \cite{Kotelnikov+2022NF_62_096025}. Parameter $R$ again varied from $R=1.1$ to $R=M$.

Critical beta values for the case $\Lambda=1$, when the conducting lateral wall is removed to infinity, have not been found. However, such values were found for $\Lambda \to 1+$.
%
%
The discussion of the stability zone at values of $\Lambda $ close to unity is relegated to Appendix \ref{A4}, since it is of academic rather than practical interest because stable zone $\beta_{\text{crit}}<\beta <1$ is extremely narrow is this case.

\begin{figure*}
  \centering
\includegraphics[width=\linewidth]{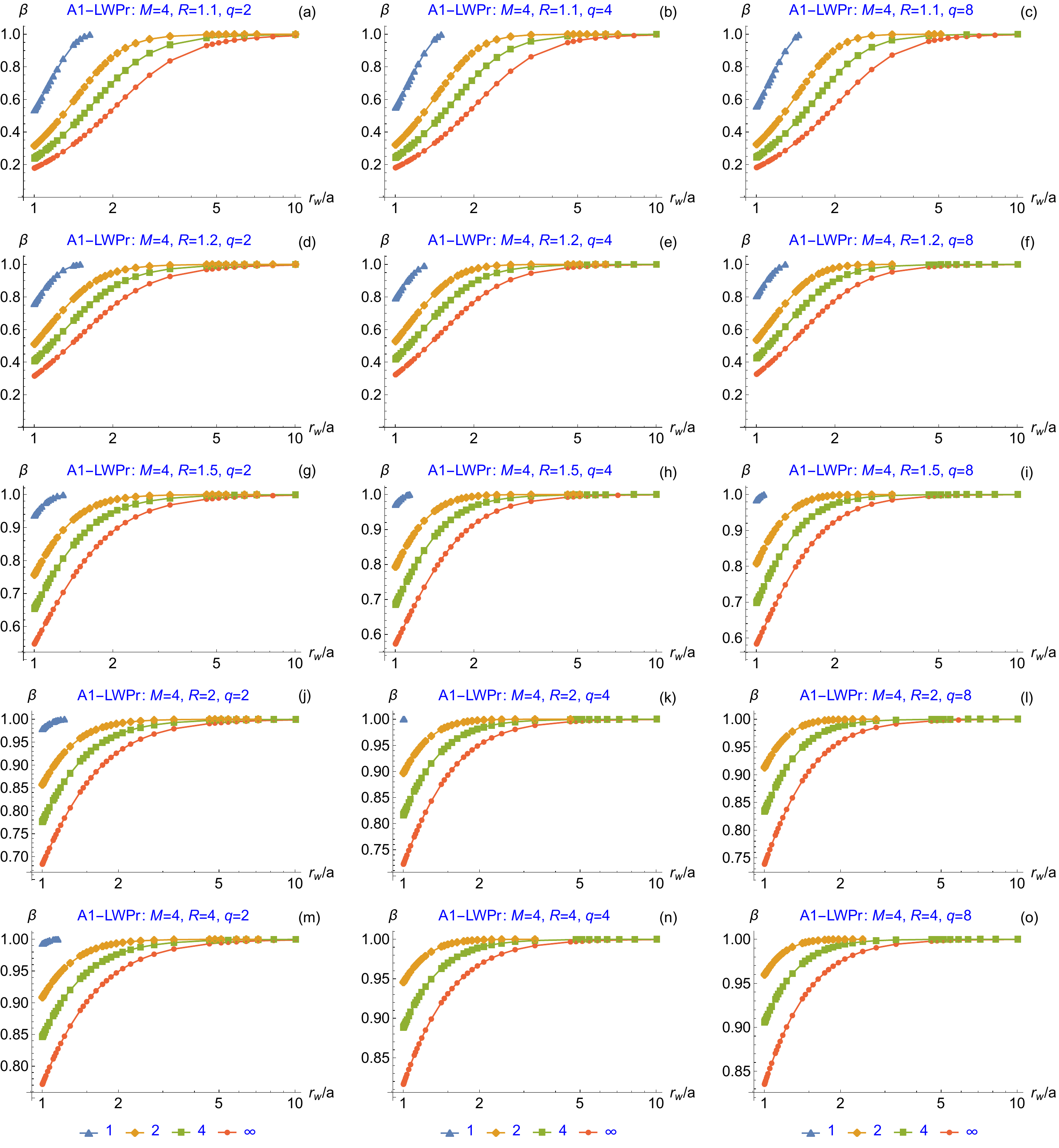}
  \caption{
      Critical beta for model magnetic field \eqref{4:35}, proportional chamber and anisotropic plasma pressure model \eqref{3:01} simulating normal NB injection;
      $M=4$, $R=\{1.1,1.2,1.5,2,4\}$.
  }\label{fig:GMA-Kesner_beta_vs_rw_K4}
\end{figure*}
\begin{figure*}
  \centering
\includegraphics[width=\linewidth]{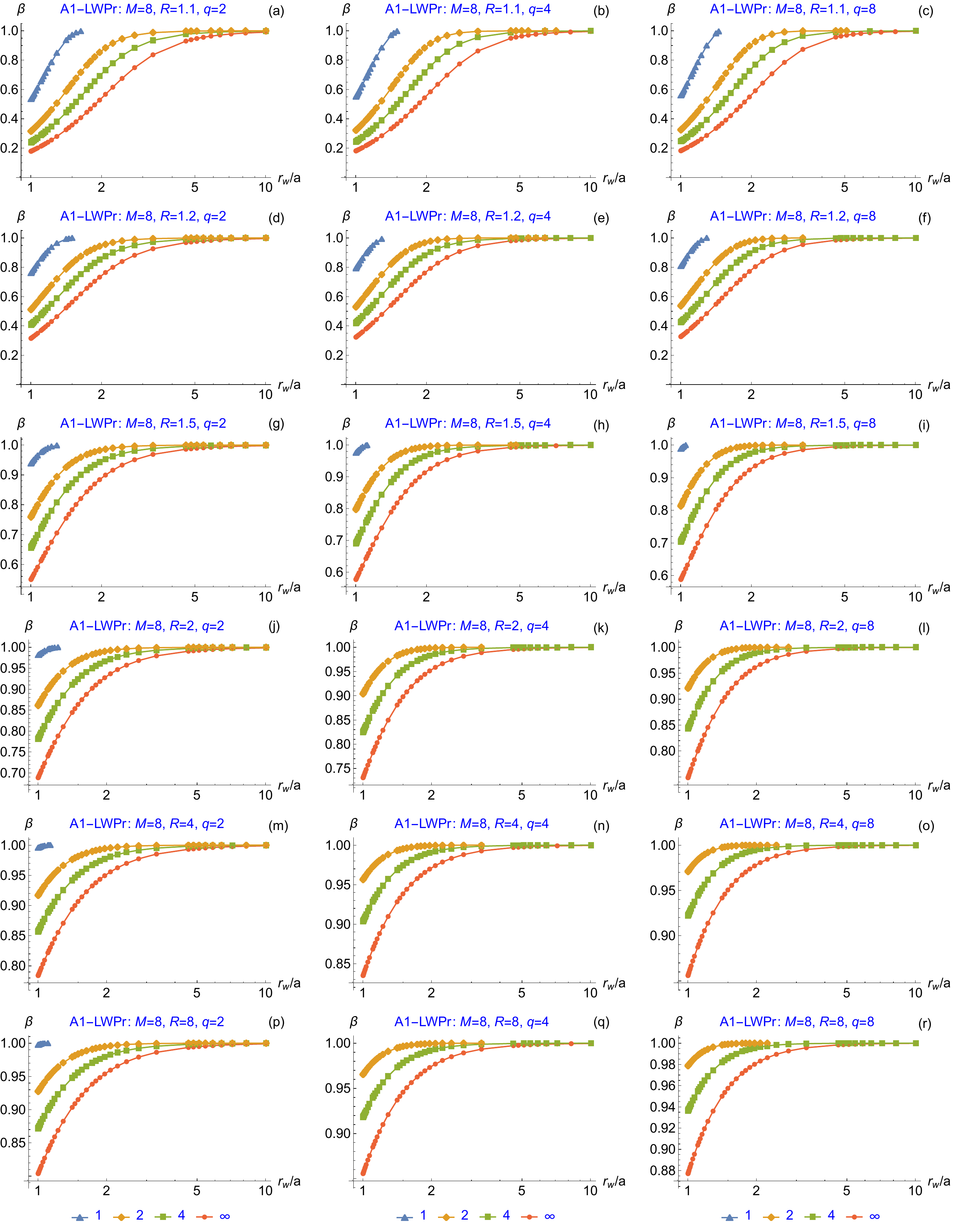}
  \caption{
      Critical beta for model magnetic field \eqref{4:35}, proportional chamber and anisotropic plasma pressure model \eqref{3:01} simulating normal NB injection;
      $ M=8$, $R=\{1.1,1.2,1.5,2,4,8\}$.
  }\label{fig:GMA-Kesner_beta_vs_rw_K8}
\end{figure*}
\begin{figure*}
  \centering
\includegraphics[width=\linewidth]{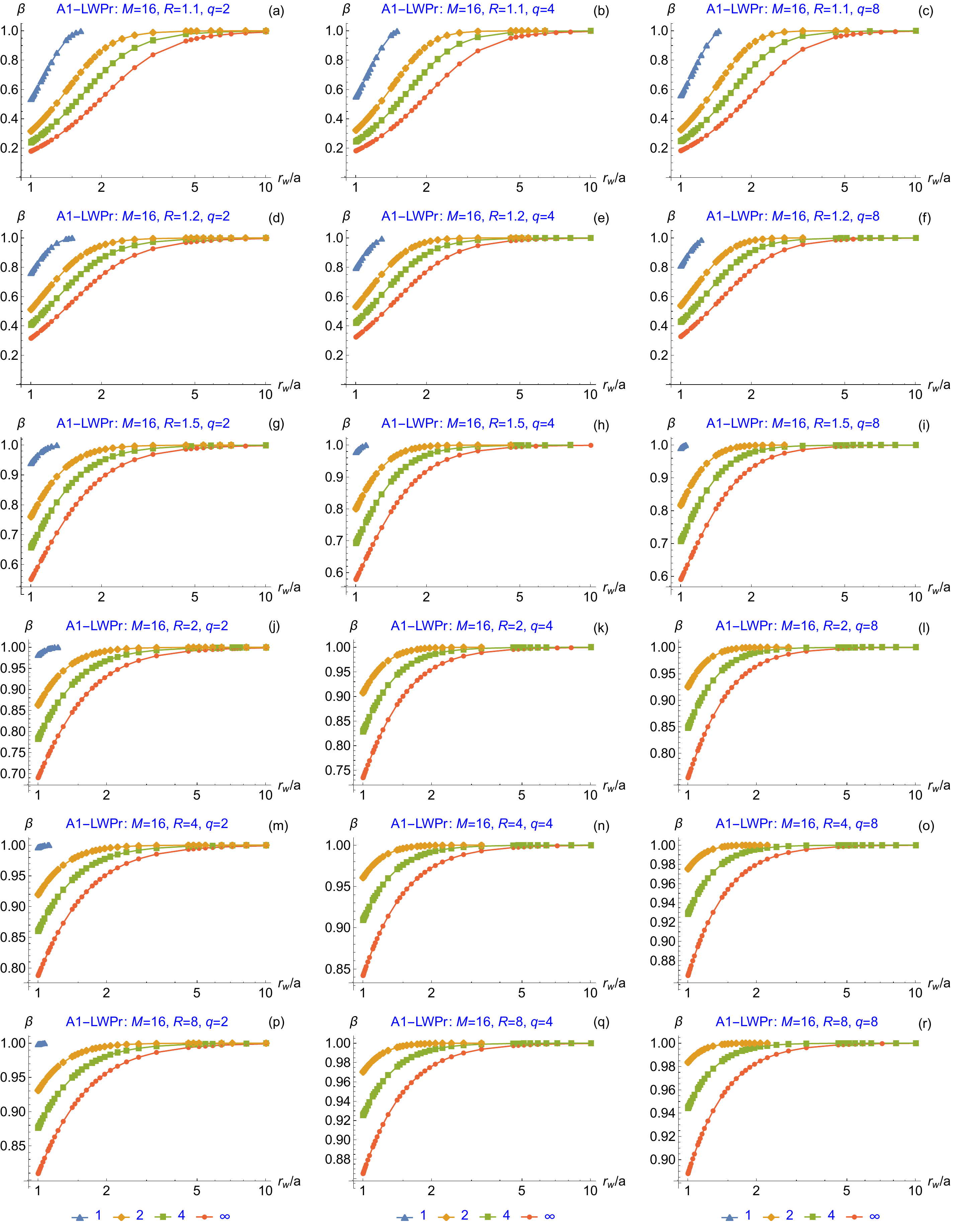}
  \caption{
      Critical beta for model magnetic field \eqref{4:35}, proportional chamber and anisotropic plasma pressure model \eqref{3:01} simulating normal NB injection;
      $M=16$, $R=\{1.1,1.2,1.5,2,4,8\}$; graphs for the case $R=16$ (not shown) are visually indistinguishable from the case $R=8$.
  }\label{fig:GMA-Kesner_beta_vs_rw_K16}
\end{figure*}

Figures \ref{fig:GMA-Kesner_beta_vs_rw_K4}, \ref{fig:GMA-Kesner_beta_vs_rw_K8}, and \ref{fig:GMA-Kesner_beta_vs_rw_K16} show plots of $\beta_{\text{crit}}$ versus  ratio $r_{w}/a=\sqrt{(\Lambda+1)/(\Lambda-1)}$ respectively for mirror ratios $4$, $8$, $16$ at various combinations of $k$, $q$, and $R$. Comparison of graphs in consecutive rows of each figure demonstrates strong dependence of critical betas on parameter $R$, which characterizes the degree of anisotropy (the smaller the $R$, the stronger the anisotropy, the lower the $\beta_{\text{crit}}$) and the spacial width of the pressure distribution (the larger the $R$, the wider the distribution).  Comparison of graphs  within each row confirms a weak tendency noted in section \ref{s4} to increase critical beta as the magnetic mirrors steepen with increasing index $q$. Comparison of figures \ref{fig:GMA-Kesner_beta_vs_rw_K4}–\ref{fig:GMA-Kesner_beta_vs_rw_K16} for different $M$ demonstrates some decrease in the stability zone as the mirror ratio increases.

We also see that the most smooth radial profile ($k=1$) is unstable for all ratios $r_{w}/a$ if $R\geq2$ and $q=4$ or $q=8$. For the case of isotropic plasma \cite{Kotelnikov+2022NF_62_096025}, the stability zone disappeared also for the next steepest profile ($k=2$) and sufficiently steep axial profiles of the magnetic field, but in the case of anisotropic plasma this profile can always be made stable by decreasing the vacuum gap (i.e.\ decreasing $r_{w}/a$).

\begin{figure*}
%
%
  \centering
\includegraphics[width=\linewidth]{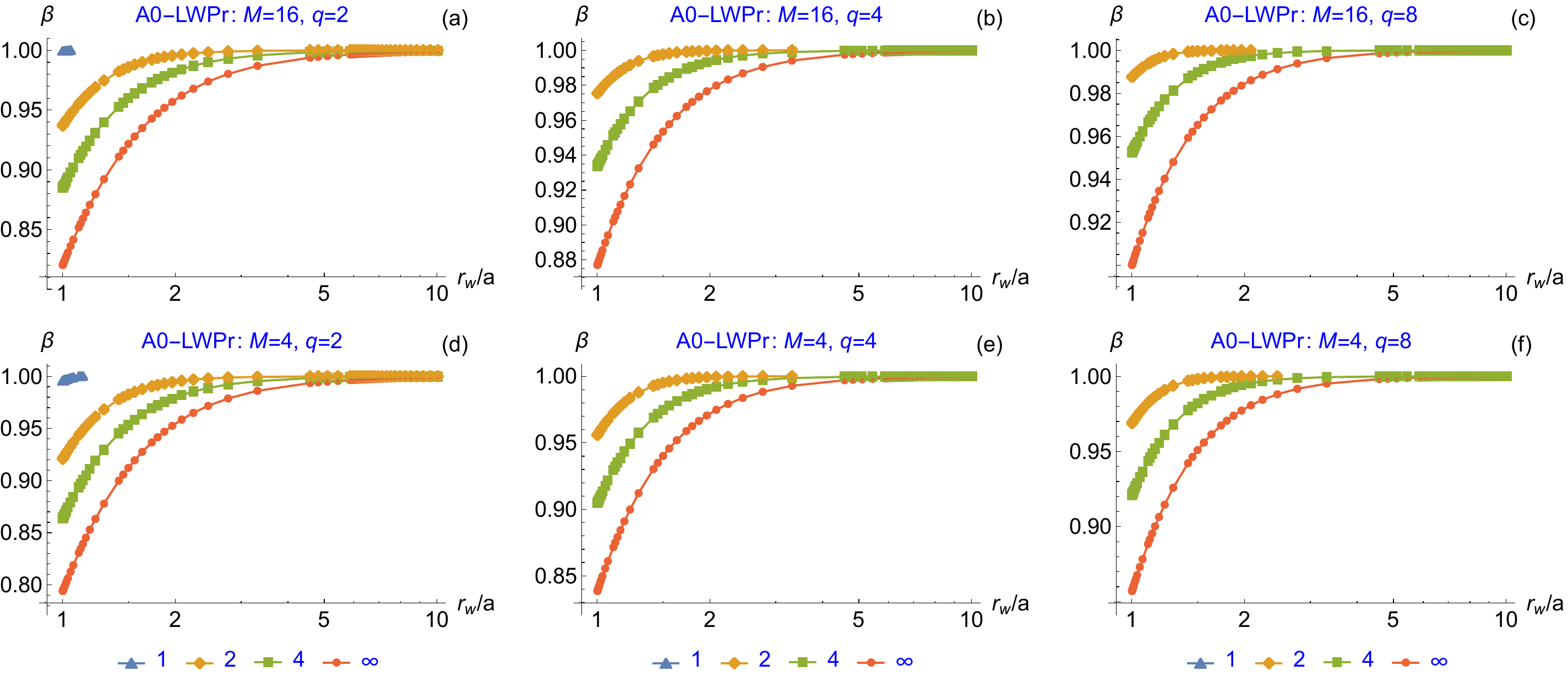}
  \\[2em]
\includegraphics[width=\linewidth]{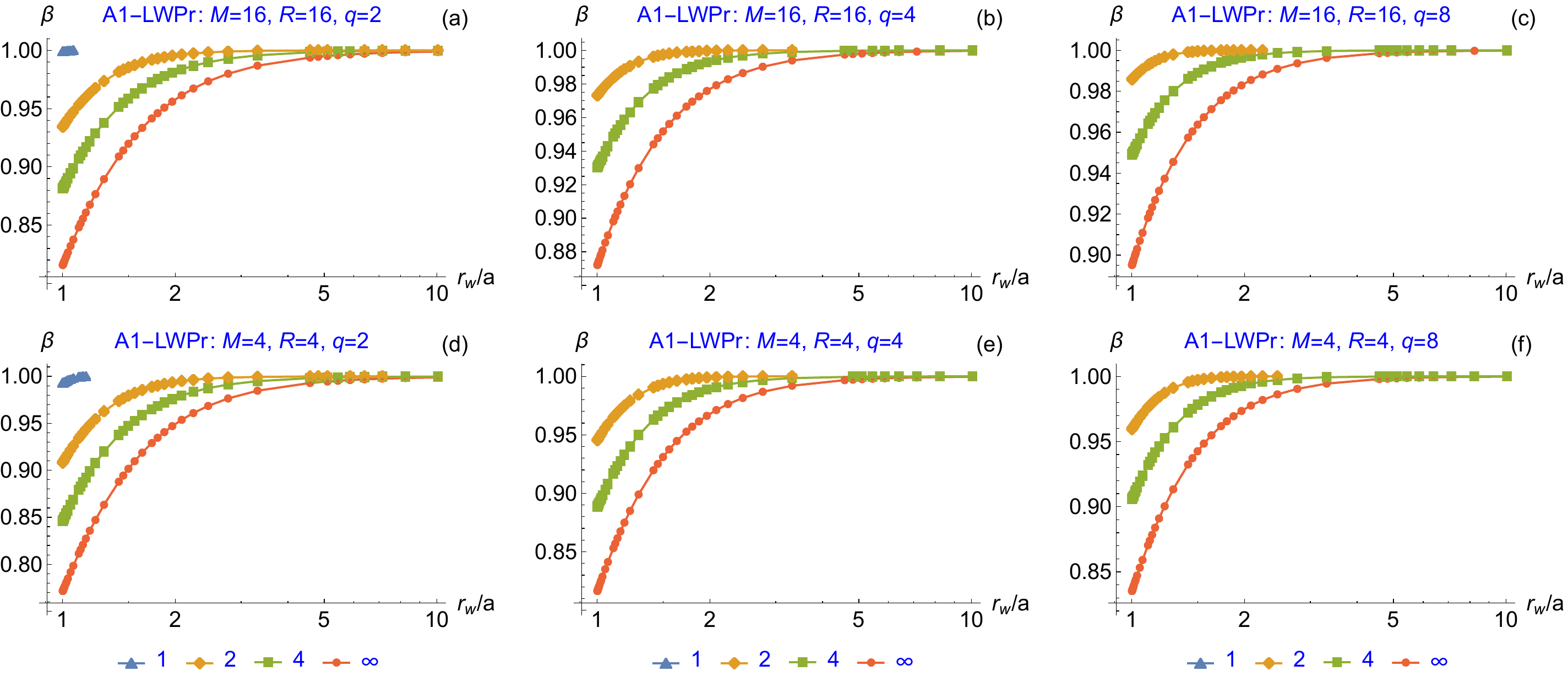}
  \caption{
    %
    Comparison of the isotropic plasma model \cite{Kotelnikov+2022NF_62_096025} with the anisotropic plasma model \eqref{3:01} with minimum anisotropy $R=M$. Graphs in odd and even rows show the critical beta for isotropic plasma and anisotropic plasma, respectively.
  }
  \label{fig:GM-Kesner-A_beta_vs_rw_R=K::}
\end{figure*}
%
For the minimum anisotropy allowed in our calculations, that is, for $R=M$, the model of normal injection of neutral beams \eqref{3:01} and the model of isotropic plasma give close results. This fact is confirmed by Fig.~\ref{fig:GM-Kesner-A_beta_vs_rw_R=K::}. The small difference becomes more noticeable for smaller $M$ and larger $q$.

\smallskip

The main trends identified by our calculations are listed below:
\setlength{\itemindent}{0em}
\setlength{\leftmargini}{1em}
\begin{itemize}
  \item
  If the stability zone can in principle exist for a given set of parameters $M$, $R$, $q$ (i.e. if a numerical solution of Eq.~\eqref{4:10} was found in section \ref{s4} for this set), then it occurs if $\Lambda$ exceeds some minimum value $\Lambda_{\min}>1$.

  \item
  This value is the smaller, the steeper the radial pressure profile (the larger the parameter $k$), the smoother the axial profile of the vacuum magnetic field (the smaller $q$), the smaller the mirror ratio $M$, and the greater the anisotropy (the smaller $R$). The stability zone narrows with increasing $\Lambda_{\min}$ and may disappear altogether for smooth pressure profiles ($k=1$).

  \item
  Critical betas for the case $\Lambda=1$, when conducting lateral wall is removed  ($r_{w}/a=\infty $), have not been found.


  \item
  The narrower the vacuum gap between the plasma and the lateral conducting wall (the larger $\Lambda$), the wider the stability zone (the smaller $\beta_{\text{crit}}$).

  \item
  The minimum $\beta_{\text{crit}}$ found for the studied set of radial and axial profiles and parameter of anisotropy $R=1.1$ is about $18\%$, which is much lower than the value $70\%$ reported for isotropic plasmas. We expect $\beta_{\text{crit}}\to0$ at $R\to1$, but such a limit does not seem achievable in a real experiment.

\end{itemize}



\section{Combined wall stabilization}\label{s6}


Finally, we repeat calculations of previous section with a minor replacement. We take boundary condition \eqref{2:11}, which means that the plasma is frozen into the conducting end plates, instead of boundary condition \eqref{2:12} describing  insulating ends of a mirror trap.

In recent paper \cite{Kotelnikov+2022NF_62_096025} we reported two critical beta values for the case of isotropic plasmas, $\beta_{\text{crit}1}$ and $\beta_{\text{crit}2}$: one at low beta due to the balancing of the ponderomotive force with the curvature drive, and one at high $\beta_{}$ due to the proximity of the conducting lateral wall which enables magnetic line bending to balance the curvature drive. Corresponding to two critical values of beta, there are two zones of stability. The first zone exists at low plasma pressure, at $0<\beta<\beta_{\text{crit}1}$, and the second one exists at high pressure, at $\beta_{\text{crit}2}<\beta <$1. These two zones merge at larger $\Lambda $ providing overall stability at any beta. In this section we extend these results to the case of anisotropic plasma produced by normal NB injection. The main difference from the isotropic model is that the degree of anisotropy provides an additional ``knob'' for controlling experimental parameters.

Let's move on to solving Eq.~\eqref{2:01} with the boundary conditions \eqref{2:11} for $z=1$, and \eqref{5:01} and \eqref{5:03} for $z=0$. As shown in section \ref{s5}, in the region $z_{s}<z<1$ the desired solution satisfies Eq.~\eqref{5:14}.
Noticing that the constant in this equation is now equal to
    $
    \left[
        \Lambda(z_{s}) + 1
    \right]
    \phi'(z_{s}^{+})
    $,
rather than zero as in section \ref{s5}, we find that the derivative $\phi'(z)$ at $z=z_{s}^{+}$ is equal to
    \begin{gather}
    \label{6:03}
    \phi'(z_{s}^{+})
    =
    \frac{- \phi(z_{s})}
    {
        \left[ \Lambda(z_{s}) + 1 \right]
        \int_{z_{s}}^{1}\frac{\dif{z}}{\Lambda(z) + 1}
    }
    .
    \end{gather}
%
Substituting $\phi(z_{s}^{+})$ into Eq.~\eqref{5:15} and taking into account that $Q({z_{s}^{+}})=0$, we find the derivative $ \phi'({z_{s}^{-}})$ on the right boundary of the interval $0<z<z_{s}$ from its inner side:
    \begin{gather}
    \label{6:04a}
    \phi'({z_{s}^{-}})
    =
    \left[
        \frac{Q({z_{s}^{-}})}{
            \Lambda(z_{s}) + 1
            \vphantom{\int_{z_{s}}^{1}}
        }
        -
        \frac{1}
        {
            \left[ \Lambda(z_{s}) + 1 \right]
            \int_{z_{s}}^{1}\frac{\dif{z}}{\Lambda(z) + 1}
        }
    \right]
    \phi(z_{s})
    .
    \end{gather}
%
This is the boundary condition that should be used instead of \eqref{5:17} in the problem of ballooning instability with a combination of wall stabilization and stabilization by conducting end plates. In the case of $\Lambda(z)=\const$ under consideration here it reduces to
    \begin{equation}
    \label{6:04}
    \phi'({z_{s}^{-}})
    =
    \left[
    \frac{
        Q({z_{s}^{-}})
    }{
        \Lambda + 1
    }
    -
    \frac{
        1
    }{
        1 - z_{s}
    }
    \right]
    \phi(z_{s})
    .
    \end{equation}

\begin{figure*}
  \includegraphics[width=\linewidth]{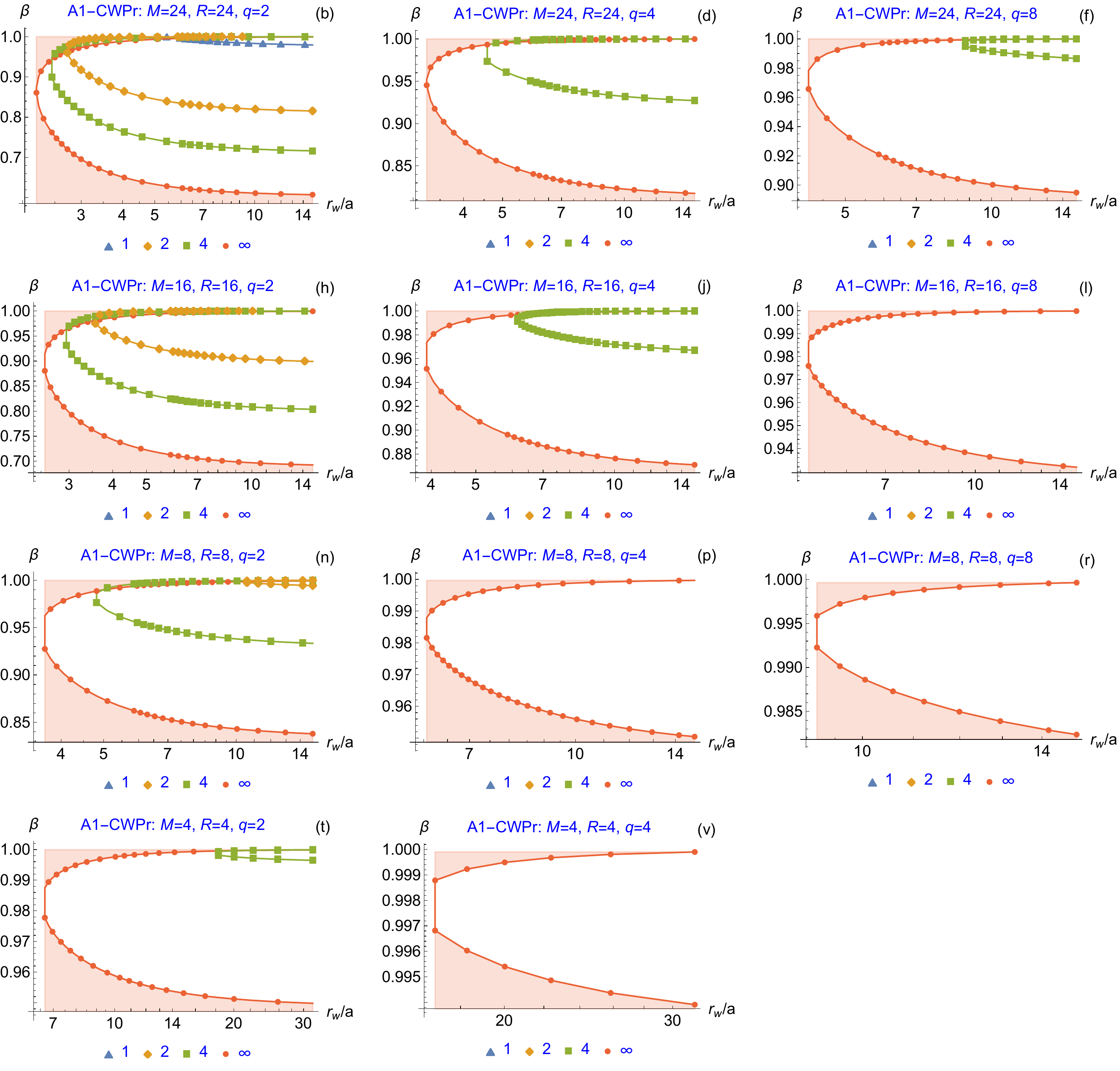}
  \caption{
    %
    Instability zone for model magnetic field \eqref{4:35},
    proportional chamber and anisotropic plasma pressure model \eqref{3:01} simulating normal NB injection at combined stabilization by end MHD anchors;
    $q=\{2,4,8\}$, $M=\{24,16,8,4\}$, minimal anisotropy $R=M$.
    %
    %
    The instability zone is located between the lower curve $\beta_{\text{crit}1}(r_{w}/a)$ and the upper curve $\beta_{\text{crit}2}(r_{w}/a )$ of one colors; the correspondence of the index $to$ to the color is shown below the figures; the instability zone is not shaded for a plasma with a sharp boundary ($k=\infty $), for which it has the maximum dimensions. Compare with fig.~\ref{fig:2022-GM-Kesner_beta_vs_rw}.  }
  \label{fig:2022-GM-Kesner-A-Ends_beta_vs_rw_R=K}
\end{figure*}


Following same scheme as in Ref.~\cite{Kotelnikov+2022NF_62_096025}, we performed a series of calculations for the vacuum magnetic field \eqref{4:35} with three values of the index $q=\{2,4,8\}$ and a mirror ratio from a limited set $M = \{24, 16, 8, 4\}$. Parameter of anisotropy $R\leq M$ was taken from the list $R=\{1.1, 1.2, 1.5, 2, 4, 8, 16, 24\}$.

A series of graphs in Fig.~\ref{fig:2022-GM-Kesner-A-Ends_beta_vs_rw_R=K} illustrates the results of calculations at the minimum degree of anisotropy, i.e. at $R=M$, when the instability zone has the maximum dimensions.
Zone of instability in this and subsequent figures for each radial pressure profile with a given index $k=\{1,2,4,\infty \}$ lies between the lower and upper curves of the corresponding color. They represent $\beta_{\text{crit}1}$ and $\beta_{\text{crit}2}$, respectively. If there is only one curve, as in Fig.~\ref{fig:2022-GM-Kesner-A-Ends_beta_vs_rw_R=K}(a) for the profile $k=1$, we interpret it as $\beta_{\text{crit} 1}$. If there are no curves of a given color on the graph, we suppose that the corresponding pressure profile is stable in the entire range of $0<\beta<1$ and in the entire range of values of $r_{w}/a$, which is shown in the figure.
%
%
We have made the last two statements, analyzing the evolution of the instability zone as the parameters of the problem, such as $r_{w}/a_{0}$, $R$, $M$, change continuously. Strictly speaking, we should have verified these conclusions by calculating the sign of $\omega^{2}$. However, our numerical code does not currently provide  this option.
%


In the range $r_{w}/a$ to the left of the left edge of the displayed interval (i.e., in the region where the ratio $r_{w}/a$ is relatively small) all radial profiles are stable over the entire interval $0<\beta<1 $. It is easy to see that smooth pressure profiles ($k=1$, $k=2$) are more stable than steep profiles ($k=4$, $k=\infty$). The same trend also takes place for small-scale ballooning perturbations when using end MHD stabilizers \cite{Kotelnikov+2021PST_24_015102}.

%
In the opposite case $r_{w}/a\to\infty$ the upper stability zone becomes extremely thin (or even disappears for smooth radial profiles), but it is located in the immediate vicinity of the $\beta=1$ boundary, where the paraxial approximation does not work. At the same time, the lower zone remains quite wide. This fact is in good agreement with the results of calculating the small-scale ballooning instability threshold \cite{BushkovaMirnov1986VANT_2_19e, RyutovStupakov1981IAEA_1_119, Kotelnikov+2021PST_24_015102}, as we emphasized in the introductory section \ref{s1}.

%
Surprise is caused by another fact, namely: the disappearance of the instability zone at $r_{w}/a\to\infty$ as the radial pressure profile is smoothed out. For the sharpest profile $k=\infty $, the instability zone is present on all graphs in Fig.~\ref{fig:2022-GM-Kesner-A-Ends_beta_vs_rw_R=K}; for the $k=4$ profile it is visible in seven out of eleven graphs, for the profile $k=2$ only on two, and for $k=1$ only on one. We interpret the absence of a gap between the lower and upper stability zones for a given ratio $r_{w}/a$ as the stability of rigid ballooning perturbations over the entire interval $0<\beta<1$. But then the conclusion is inevitable that a sufficiently smooth radial pressure profile can be stable in the absence of a lateral wall for any value of $\beta$. Note that the absence of graphs for the case $M=2$, $q=\{4,8\}$ in Fig.~\ref{fig:2022-GM-Kesner-A-Ends_beta_vs_rw_R=K} means that there is no instability zone even for the sharpest radial profile $k=\infty $.

\begin{figure*}
\centering
\includegraphics[width=\linewidth]{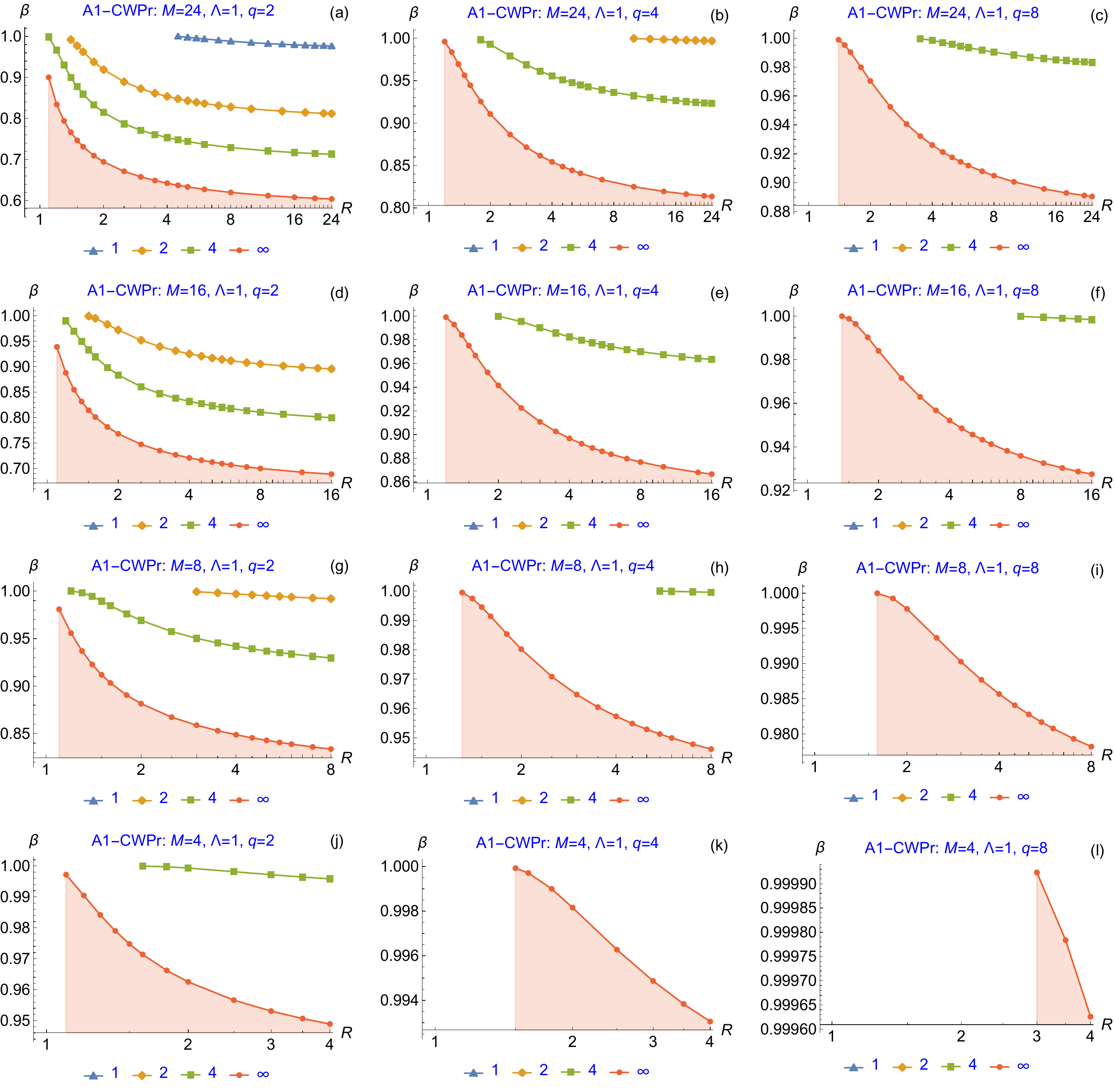}
\caption{
    %
    Boundary of lower stability zone $\beta_{\text{crit}1}$ for model magnetic field \eqref{4:35} and anisotropic plasma model \eqref{3:01} simulating normal injection of neutral beams, with stabilization only by end conducting plates without lateral conductive wall. Stability zone for the radial profile with index $k$ is located below the corresponding curve. The absence of a curve means that the corresponding profile is stable over the entire interval $0<\beta<1$.
}
\label{fig:GM2-A1-CWPr_beta_vs_R-minL}
\end{figure*}

\begin{figure*}
  \centering
%
\includegraphics[width=\linewidth]{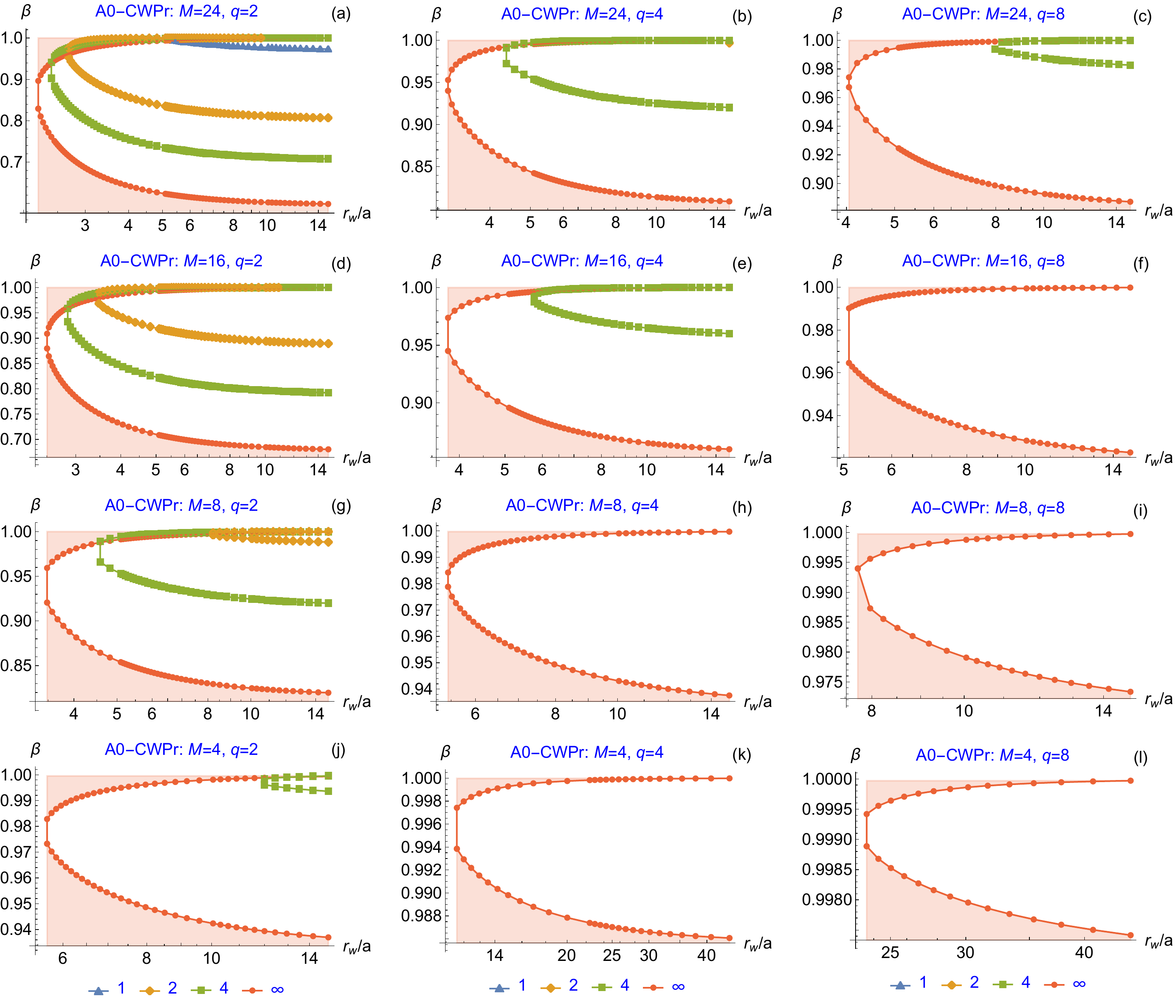}
  \caption{
    Stability zone for model magnetic field \eqref{4:35},
    proportional chamber and isotropic plasma at combined stabilization by end MHD stabilizers;
    $q=\{2,4,8\}$, $M=\{24,16,8,4\}$.
    Instability zone is located between $\beta_{\text{crit}1}(r_{w}/a)$ (lower curve) and $\beta_{\text{crit}2}(r_{w}/a)$ (upper curve of the same color). The instability zone is not shaded for a plasma with a sharp boundary ($k=\infty$), for which it has maximum dimensions. Compare with Fig.~\ref{fig:2022-GM-Kesner-A-Ends_beta_vs_rw_R=K}.
  }\label{fig:2022-GM-Kesner_beta_vs_rw}
\end{figure*}
For comparison, Fig.~\ref{fig:2022-GM-Kesner_beta_vs_rw} shows the results of calculations for isotropic plasma with the same combinations of indices $q$ and mirror ratios $M$ as in Fig.~\ref{fig:2022-GM-Kesner-A-Ends_beta_vs_rw_R=K}. As can be seen from these two figures, \ref{fig:2022-GM-Kesner-A-Ends_beta_vs_rw_R=K} and~\ref{fig:2022-GM-Kesner_beta_vs_rw}, the dimensions of the instability zones for the same mirror ratio are very close, but in the case of an anisotropic plasma they are still smaller, which becomes more noticeable as the mirror ratio decreases. Both in the case of anisotropic and isotropic plasma models, the dimensions of the instability zone $\beta_{\text{crit}1} < \beta < \beta_{\text{crit}2}$ (which is not shaded) are maximum for the smoothest magnetic field profile with index $q=2$. They are minimal at $q=8$.

%
On the contrary, as can be seen from the analysis of figures \ref{fig:GMA-Kesner_beta_vs_rw_K4}, \ref{fig:GMA-Kesner_beta_vs_rw_K8} and \ref{fig:GMA-Kesner_beta_vs_rw_K16} in section \ref{s5}, with only wall stabilization implemented without end plates installed, the instability zone at $\beta < \beta _{\text{crit}}$ gets slightly larger as $q$ increases. In case of combined stabilization, for a fixed set of parameters $q$, $M$, $R$ the instability zone is maximum for the steepest radial pressure profile ($k=\infty$) and may be completely absent for smooth profiles ($k=1 $, $k=2$). Without stabilization by the conducting end plates, the opposite situation takes place: the stability zone is smaller and may be completely absent for smooth profiles. It means that the effect of end plates dominates. Another important effect of end MHD stabilizers is a much stronger dependence on the mirror ratio $M$ and the axial profile of the magnetic field characterized by the index $q$.


\begin{figure*}
  \centering
%
\includegraphics[width=\linewidth]{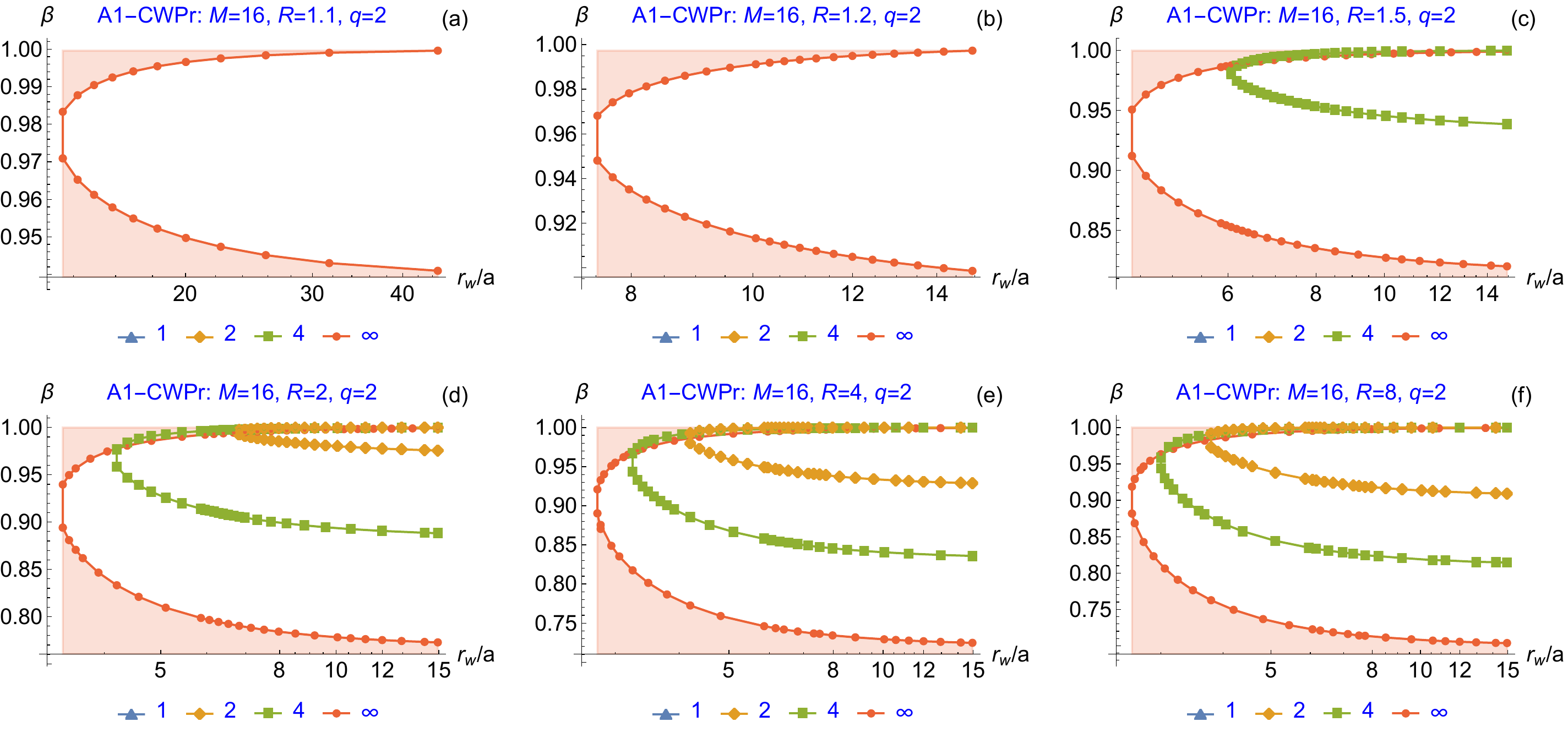}
\caption{
    Stability zone versus ratio $r_{w}/a$ for model magnetic field \eqref{4:35},
    proportional conducting chamber and anisotropic plasma pressure model \eqref{3:01} simulating normal NB injection at combined stabilization by end MHD stabilizers;
    $q=\{2\}$, $M=\{16\}$, various anisotropy $R=\{1.1, 1.2, 1.5, 2,4,8\}$.
    %
    %
    Instability zone is located between $\beta_{\text{crit}1}(r_{w}/a)$ (lower curve) and $\beta_{\text{crit}2}(r_{w}/a)$ (upper curve of the same color); it is not shaded for a plasma with a sharp boundary ($k=\infty $), for which it has the maximum dimensions.
  }
  \label{fig:2022-GM-Kesner-A-Ends_beta_vs_rw_q2_K16}
\end{figure*}
%

Increasing the anisotropy only enhances the effect of the end plates. A series of graphs in Fig.~\ref{fig:2022-GM-Kesner-A-Ends_beta_vs_rw_q2_K16} illustrates the dependence of the stability/instability zones on the anisotropy parameter $R$ for a fixed index $q=2$ and a fixed mirror ratio $M=16$. The unshaded zone of instability noticeably decreases with increasing anisotropy (with decreasing parameter $R$) and may even disappear, as for $R=1.1$.

\medskip

Summing up everything said in this section, we can state the following:
\setlength{\itemindent}{0em}
\setlength{\leftmargini}{1em}
\begin{itemize}
  \item
    %
    Two stability zones are found for moderate values of parameter $\Lambda$ and a sufficiently large mirror ratio $M$. The lower zone $\beta < \beta_{\text{crit}1}$ exists even for $\Lambda =1$.

  \item
    %
    With other things being equal, the instability zone is maximum for the steepest pressure profile ($k=\infty $) and might not exist at all for smooth radial profiles ($k=1$ or $k=2$). Recall that in section \ref{s5} it is for the stepwise  profile that the instability zone had the minimum dimensions.

  \item
    %
    For a fixed value of the parameter $\Lambda $, the stability zones expand and can merge with a decrease in the mirror ratio $M$ and/or a smoothing of the radial pressure profile (with a decrease in $k$).

  \item
    %
    Zone of instability decreases and may even disappear as the plasma anisotropy increases (as the parameter $R$ decreases). The dimensions of the unstable zone are maximum at the minimum studied degree of anisotropy $R=M$, considered for a given mirror ratio $M$. In this limit, the dimensions of the unstable zone are close to those found for an isotropic plasma.

  \item
    %
    If an instability zone exists between two stability zones for some combinations of the parameters $k$, $q$, $M$, and $R$, then it disappears if $\Lambda > \Lambda_{\text{crit}}>1$. The value of $\Lambda_{\text{crit}}$ is the smaller, the smaller $k$, $M$ and the larger $q$. The largest value $\Lambda_{\text{crit}}=1.52$ ($r_{w}/a=\num{2.201398157116028}$) in our calculations was found at $q=2$, $M=R= $24, $k=\infty $.

%

  \item
    %
    With a not too large mirror ratio, and/or a sufficiently high plasma anisotropy, and/or a sufficiently steep magnetic field, and/or a sufficiently smooth pressure profile, and/or sufficiently close lateral wall, the rigid ballooning mode $m=1$ can be stabilized at any feasible value of beta.

\end{itemize}


\section{Conclusions}\label{s9}

In the present work, feasibility of stabilization of the $m=1$ rigid flute and ballooning modes in an axially symmetric mirror trap by specially shaped conducting cylindrical wall that repeats the shape of the plasma column on an enlarged scale is investigated using the anisotropic pressure model \eqref{3:01}, which simulates normal injection of fast neutral beams. Unlike most predecessors, who studied not quite realistic stepwise radial plasma pressure profile, we considered four variants of the diffuse pressure profile \eqref{3:22} with different degrees of steepness, specified by the index $k$, as well as several variants of the axial profile vacuum magnetic field given by the function \eqref{4:35} with different values of index $q$ for different mirror ratios $M$.

Stabilization by a conducting wall without any additional means of MHD stabilisation is achieved at a sufficiently high plasma pressure, when the parameter $\beta$ (the dimensionless ratio of the plasma pressure to the magnetic field pressure) exceeds a certain critical value $\beta_{\text{crit}}$. Therefore, our goal was to calculate this critical value and study its dependence on the degree of plasma anisotropy, the shape of the radial pressure profile, the axial profile of the magnetic field, the mirror ratio, and the size of the vacuum gap between the plasma and the conducting wall. For calculations, we developed a numerical code, which solved Eq.~\eqref{2:01}, previously derived by Lynda LoDestro.

Our calculations showed that the stability zone expands significantly due to a decrease in the critical beta $\beta_{\text{crit}}$ as the degree of plasma anisotropy increases. The mirror ratio and axial profile of the magnetic field have relatively smaller effect on the value of $\beta_{\text{crit}}$. The dependence of $\beta_{\text{crit}}$ on the radial profile and the gap width between the plasma column and the lateral conducting wall are more significant.

From a practical point of view, a noticeable decrease in the value of $\beta_{\text{crit}}$ is achieved if the radius of the conducting wall $r_{w}$ exceeds the plasma radius $a$ by no more than $2$ times, $r_{w}/a \leq2$. The influence of the conducting wall practically disappears (in the sense that $\beta_{\text{crit}}\to1$) if $r_{w}/a\geq4$. For effective wall stabilization, the ratio $r_{w}/a$ must be less than the indicated limits. On the other hand, too small a vacuum gap between the plasma and the wall worsens the vacuum conditions due to the recycling of neutrals into the plasma from the wall surface, which leads to degradation of the plasma parameters. For this reason, plasma stabilization using only the lateral wall might seem difficult to implement, but the parameter range near $r_{w}/a=2$ seems quite comfortable from the point of view of many experimenters.

Even more promising is the stabilization of the rigid ballooning mode with a combination of a conducting lateral wall and conducting end plates, which imitate the attachment of end MHD anchors to the central cell of a mirror trap. In contrast to pure wall stabilization, mirror ratio and magnetic field profile have strong effect of the combined stabilization.

Our calculations have shown the great efficiency of this method of stabilization. We found the existence of two stability zones. One at low beta due to the balancing of the lateral wall effect with the curvature drive, and one at high $\beta_{}$ due to the proximity of the conducting wall which enables magnetic line bending to balance the curvature drive. Corresponding to two critical values of beta, there are two zones of stability. The first zone exists at low plasma pressure, at $0<\beta<\beta_{\text{crit}1}$, and the second one exists at high pressure, at $\beta_{\text{crit}2}<\beta <$1. These two zones merge, making the entire range of allowable beta values $0<\beta<1$ stable, as the mirror ratio decreases, as the vacuum gap between the plasma and the lateral wall decreases, or as the plasma pressure anisotropy increases.



\begin{acknowledgements}


    This work
    was supported by Chinese Academy of Sciences President’s International Fellowship Initiative (PIFI) under the Grants No.~2022VKA0007, No.~2022VKB0001, No.~2021VKB0013, and Chinese Academy of Sciences International Partnership Program under the Grant No.~116134KYSB20200001.

    The authors are grateful to Alexei Beklemishev for discussing the applicability conditions for the LoDestro equation.


\end{acknowledgements}

\section*{ORCID iDs}


\noindent
Igor KOTELNIKOV \href{https://orcid.org/0000-0002-5509-3174}{https://orcid.org/0000-0002-5509-3174}
\\
Vadim PRIKHODKO \href{https://orcid.org/0000-0003-0199-3035}{https://orcid.org/0000-0003-0199-3035}
\\
Dmitri YAKOVLEV \href{https://orcid.org/0000-0002-2224-4618}{https://orcid.org/0000-0002-2224-4618}

\appendix
\begin{widetext}
\section{Computing $a^{2}_{k}$}\label{A1}

Indicating the values of $k$ as a subscript, we write down the result of calculating the integral in Eq.~\eqref{2:03} by Wolfram \emph{Mathematica}$^{\copyright}$:
    \begin{subequations}
    \label{A1:1}
    \begin{equation}
    \label{A1:31-1}
    a^{2}_{1}
    =
    \frac{1}{B_{s} p_{0}}
    \left\{
    {
    B_{s} B_{v}
    -
    \sqrt{\left(B_{s}^2-2p_{0}\right)
    \left(B_{v}^2-2p_{0}\right)}
    +
    \left(B_{s}^2-B_{v}^2\right)
    \left[
    \ln\left(
        \frac{B_{s}+B_{v}}
        {
            \sqrt{B_{s}^{2}-2p_{0}}+\sqrt{B_{v}^{2}-2p_{0}}
        }
    \right)
    \right]
    }
    \right\}
    ;
\end{equation}
    \begin{equation}
    \label{A1:32f}
    a_{2}^{2}=
    \frac{\sqrt{\frac{2 B_{s}^2}{p_{0}}-4}
    }{B_{s}}
    \left[
        F\left(
            \tan^{-1}\left(
                \sqrt{\frac{2p_{0}}{B_{v}^2-2 p_{0}}}
            \right)
            \Biggm|
            \frac{B_{s}^2-B_{v}^2}{B_{s}^2-2 p_{0}}
        \right)
        -
        E\left(
            \tan ^{-1}\left(
                \sqrt{\frac{2p_{0}}{B_{v}^2-2 p_{0}}}
            \right)
            \Biggm|
            \frac{B_{s}^2-B_{v}^2}{B_{s}^2-2 p_{0}}
        \right)
        +
        \frac{B_{s}}{B_{v}}
        \sqrt{\frac{2p_{0}}{B_{s}^2-2 p_{0}}}
    \right]
    ;
    \end{equation}

    \begin{equation}
    \label{A1:33f}
    a^{2}_{4}
    =
    \frac{2}{B_{s}}
    \sqrt{\frac{B_{s}^2-2p_{0}}{B_{v}^2-2p_{0}}}
    \,
    F_1\left(
            \frac{1}{4};
            -\frac{1}{2},
            \frac{1}{2};\frac{5}{4};
            -\frac{2p_{0}}{B_{s}^2-2p_{0}},
            -\frac{2p_{0}}{B_{v}^2-2p_{0}}
    \right)
    ;
    \end{equation}


    \begin{equation}
    \label{A1:34f}
    a^{2}_{\infty }
    =
    \frac{2}{B_{s}}
    \sqrt{\frac{B_{s}^2-2p_{0}}{B_{v}^2-2p_{0}}}
    .
    \end{equation}
    \end{subequations}
Here $F(\phi | m)$ and $E(\phi | m)$ are the elliptic integral of the first and the second kinds, respectively., and $F_1(a; b_{1},b_{2}; c; x,y)$ is the Appell hypergeometric function of two variables.



\section{Integrals $P_{k}$}\label{A2}

Indicating again the values of $k$ as a subscript, we write down the result of calculating the integral in Eq.~\eqref{2:06} by Wolfram \emph{Mathematica}$^{\copyright}$:
    \begin{subequations}
    \label{3:41}
    \begin{multline}
    \label{3:41a}
    \frac{a_{1}^{2}}{2}\mean{p_{\bot}}_{1}
    =
    -\frac{1}{
        16 B_{s} p_{0}
    }\,
    \left\{
        {
        B_{s}^3 B_{v}
        -
        3 B_{s} B_{v}^3
        +
        8 p_{0}^2
        +
        \left(
        -
        B_{s}^2
        +
        4 p_{0}
        +
        3 B_{v}^2
        \right)
        \sqrt{
            \left(B_{s}^2-2p_{0}\right)
            \left(B_{v}^2-2p_{0}\right)
        }
    }\right.
    \\
    \left.{
        +
        \left(
            B_{s}^4
            +
            2B_{s}^2 B_{v}^2
            -
            3 B_{v}^4
        \right)
        \left[
    \ln\left(
        \frac{
            \sqrt{B_{s}^{2}-2p_{0}}+\sqrt{B_{v}^{2}-2p_{0}}
        }{B_{s}+B_{v}}
    \right)
        \right]
    }
    \right\}
    .
    \end{multline}
    \begin{multline}
    \label{3:42f}
    \frac{a_{2}^{2}}{2}\mean{p_{\bot}}_{2}
    =
    \frac{1}{12 B_{s}^2}
    \left\{
        \frac{1}
        {
            \sqrt{p_{0}  \left(B_{v}^2-2 p_{0}\right)}
        }
        \left[
            \sqrt{2} B_{s}
            \left(
                \left(B_{s}^2-2 p_{0}\right)
                \left(B_{v}^2+4 p_{0}\right)
                F\left(
                    \tan^{-1}\left(
                        \sqrt{\frac{2p_{0}}{B_{s}^2-2p_{0}}}
                    \right)
                    \Biggm|
                    \frac{B_{v}^2-B_{s}^2}{B_{v}^2-2 p_{0}}
                \right)
            \right.
        \right.
    \right.
    \\
    \left.
        \left.
            \left.
                +
                \left(B_{v}^2-2 p_{0}\right)
                \left(
                    B_{s}^2-2 \left(B_{v}^2+2p_{0}\right)
                \right)
                E\left(
                    \tan^{-1}\left(
                        \sqrt{\frac{2p_{0}}{B_{s}^2-2p_{0}}}
                    \right)
                    \Biggm|
                    \frac{B_{v}^2-B_{s}^2}{B_{v}^2-2 p_{0}}
                \right)
            \right)
        \right]
    \right.
    \\
    \left.
        -
        4 (B_{s}-B_{v})(B_{v} (B_{s}+B_{v})+2 p_{0})
    \right\}
    ;
    \end{multline}

    \begin{multline}
    \label{3:43f}
    \frac{a_{4}^{2}}{2}\mean{p_{\bot}}_{4}
    =
    \left\{
        5 \left(B_{s}^2-2 p_{0}\right)
        \left(B_{v}^2+8 p_{0}\right)
        F_1\left(
            \frac{1}{4};
            \frac{1}{2},
            \frac{1}{2};
            \frac{5}{4};
            -\frac{2 p_{0}}{B_{s}^2-2 p_{0}},
            -\frac{2 p_{0}}{B_{v}^2-2 p_{0}}
        \right)
        +
    \right.
    \\
    \left.
        +
        2 p_{0} \left(-2 B_{s}^2+3 B_{v}^2+8 p_{0}\right)
        F_1\left(
            \frac{5}{4};
            \frac{1}{2},
            \frac{1}{2};
            \frac{9}{4};
            -\frac{2 p_{0}}{B_{s}^2-2 p_{0}},
            -\frac{2 p_{0}}{B_{v}^2-2 p_{0}}
        \right)
        -
    \right.
    \\
        -
    \left.
        5 \sqrt{
            \left(B_{s}^2-2 p_{0}\right)
            \left(B_{v}^2-2 p_{0}\right)
        }
        \left(B_{s} B_{v}+8 p_{0}\right)
    \right\}
    \Bigm/
    \left[
        50 B_{s}
        \sqrt{\left(B_{s}^2-2 p_{0}\right) \left(B_{v}^2-2 p_{0}\right)}
    \right]
    ;
    \end{multline}
    \begin{equation}
    \label{3:44f}
    \frac{a_{\infty}^{2}}{2}\mean{p_{\bot}}_{\infty}
    =
    \frac{p_{0}}{B_{s}}
    \left(
        \sqrt{\frac{B_{s}^2-2p_{0}}{B_{v}^2-2p_{0}}}
        -1
    \right)
    .
   \end{equation}
   \end{subequations}
Here $F(\phi | m)$ and $E(\phi | m)$ are the elliptic integral of the first and the second kinds, respectively., and $F_1(a; b_{1},b_{2}; c; x,y)$ is the Appell hypergeometric function of two variables.

\end{widetext}

\section{Minimal \texorpdfstring{$\Lambda$}{Lambdas}}\label{A4}

\begin{table*}[!htp]
  \centering
   \begin{equation*}
  \begin{array}{|c|c|cccccccccc|}
\hline
  \multicolumn{12}{|c|}{K=16,\qquad \beta=0.999}\\
\hline
 k & q\backslash R & 1.1 & 1.2 & 1.5 & 2 & 3 & 4 & 6 & 8 & 12 & 16
 \\
\hline
 \text{} & 2 & 2.13 & 2.62 & 4.01 & 6.45 & 11.8 & 17.7 & 30.3 & 44.0 & 75.4 & 113.
 \\
 1 & 4 & 2.55 & 3.61 & 10.5 & \text{N/F} & \text{N/F} & \text{N/F} & \text{N/F} & \text{N/F} & \text{N/F} & \text{N/F}
 \\
 \text{} & 8 & 2.74 & 4.16 & 23.9 & \text{N/F} & \text{N/F} & \text{N/F} & \text{N/F} & \text{N/F} & \text{N/F} & \text{N/F}
 \\
\hline
 \text{} & 2 & 1.069 & 1.10 & 1.17 & 1.22 & 1.26 & 1.27 & 1.29 & 1.29 & 1.30 & 1.30
 \\
 2 & 4 & 1.15 & 1.23 & 1.41 & 1.59 & 1.77 & 1.86 & 1.94 & 1.98 & 2.02 & 2.04
 \\
 \text{} & 8 & 1.19 & 1.31 & 1.59 & 1.92 & 2.34 & 2.58 & 2.83 & 2.96& 3.08 & 3.15
 \\
\hline
 \text{} & 2 & 1.013 & 1.022 & 1.038 & 1.051 & 1.061 & 1.065 & 1.068 & 1.069 & 1.070 & 1.070
 \\
 4 & 4 & 1.030 & 1.052 & 1.095 & 1.13 & 1.17 & 1.18 & 1.19 & 1.20 & 1.20 & 1.20
 \\
 \text{} & 8 & 1.044 & 1.076 & 1.14 & 1.21 & 1.27 & 1.30 & 1.32 & 1.33 & 1.34 & 1.35
 \\
\hline
 \text{} & 2 & 1.0033 & 1.0059 & 1.011 & 1.015 & 1.018 & 1.019 & 1.019 & 1.020 & 1.020 & 1.020
 \\
 \infty  & 4 & 1.0073 & 1.013 & 1.025 & 1.034 & 1.042 & 1.045 & 1.048 & 1.049 & 1.049 & 1.050
 \\
 \text{} & 8 & 1.011 & 1.019 & 1.037 & 1.053 & 1.066 & 1.071 & 1.075 & 1.077 & 1.079 & 1.079
 \\
\hline
\end{array}
  \end{equation*}
  \begin{equation*}
  \begin{array}{|c|c|cccccccc|}
\hline
  \multicolumn{10}{|c|}{K=8,\qquad \beta=0.999}\\
\hline
 k & q\backslash R & 1.1 & 1.2 & 1.5 & 2 & 3 & 4 & 6 & 8 \\
\hline
 \text{} & 2 & 2.13 & 2.61 & 3.97 & 6.27 & 10.8 & 15.0 & 21.6 & 26.4 \\
 1 & 4 & 2.54 & 3.58 & 9.88 & \text{N/F} & \text{N/F} & \text{N/F} & \text{N/F} & \text{N/F}
 \\
 \text{} & 8 & 2.72 & 4.10 & 20.2 & \text{N/F} & \text{N/F} & \text{N/F} & \text{N/F} & \text{N/F}
 \\
\hline
 \text{} & 2 & 1.069 & 1.10 & 1.17 & 1.22 & 1.26 & 1.27 & 1.28 & 1.29 \\
 2 & 4 & 1.15 & 1.23 & 1.40 & 1.57 & 1.73 & 1.80 & 1.87 & 1.90
 \\
 \text{} & 8 & 1.19 & 1.31 & 1.57 & 1.87 & 2.21 & 2.39 & 2.57 & 2.64 \\
\hline
 \text{} & 2 & 1.013 & 1.022 & 1.038 & 1.051 & 1.061 & 1.064 & 1.067 & 1.068
 \\
 4 & 4 & 1.030 & 1.051 & 1.093 & 1.13 & 1.16 & 1.17 & 1.18 & 1.19 \\
 \text{} & 8 & 1.043 & 1.074 & 1.14 & 1.20 & 1.25 & 1.28 & 1.29 & 1.30
 \\
\hline
 \text{} & 2 & 1.0033 & 1.0059 & 1.011 & 1.015 & 1.018 & 1.019 & 1.019 & 1.020
 \\
 \infty  & 4 & 1.0073 & 1.013 & 1.024 & 1.034 & 1.041 & 1.044 & 1.046 & 1.047
 \\
 \text{} & 8 & 1.010 & 1.019 & 1.036 & 1.051 & 1.063 & 1.067 & 1.071 & 1.072
 \\
\hline
\end{array}
\quad
\begin{array}{|c|c|cccccc|}
\hline
  \multicolumn{8}{|c|}{K=4,\qquad \beta=0.999}\\
\hline
 k & q\backslash R & 1.1 & 1.2 & 1.5 & 2 & 3 & 4 \\
\hline
 \text{} & 2 & 2.12 & 2.60 & 3.89 & 5.84 & 8.72 & 10.2 \\
 1 & 4 & 2.53 & 3.52 & 8.88 & \text{N/F} & \text{N/F} & \text{N/F} \\
 \text{} & 8 & 2.70 & 4.00 & 15.9 & \text{N/F} & \text{N/F} & \text{N/F} \\
\hline
 \text{} & 2 & 1.069 & 1.10 & 1.17 & 1.21 & 1.25 & 1.26 \\
 2 & 4 & 1.14 & 1.22 & 1.38 & 1.53 & 1.65 & 1.70 \\
 \text{} & 8 & 1.19 & 1.30 & 1.53 & 1.78 & 2.02 & 2.11 \\
\hline
 \text{} & 2 & 1.013 & 1.022 & 1.038 & 1.050 & 1.059 & 1.062 \\
 4 & 4 & 1.030 & 1.050 & 1.090 & 1.12 & 1.15 & 1.16 \\
 \text{} & 8 & 1.042 & 1.072 & 1.13 & 1.18 & 1.23 & 1.24 \\
\hline
 \text{} & 2 & 1.0033 & 1.0059 & 1.011 & 1.014 & 1.017 & 1.018 \\
 \infty  & 4 & 1.0072 & 1.013 & 1.024 & 1.032 & 1.039 & 1.041 \\
 \text{} & 8 & 1.010 & 1.018 & 1.034 & 1.047 & 1.058 & 1.061 \\
\hline
\end{array}
  \end{equation*}
  \caption{
    Minimal $\Lambda_{\min} $ for the magnetic field \eqref{4:35} and anisotropic pressure model \eqref{3:01}.
  }\label{tbl:minLambda}
\end{table*}



The existence of a stability zone, even if in the limit of $\beta \to 1$, at a very large gap between the conducting wall and the plasma, was previously discovered for the case of isotropic plasma \cite{Kotelnikov+2022NF_62_096025}, and then caused surprise. For $\Lambda =1.01$, the radius of the conducting wall $r_{w}$ is $14$ times larger than the plasma radius $a$. In general, the minimum $\Lambda_{\min}$, for which the critical values of beta were found, was the smaller, the steeper the radial pressure profile (the larger the parameter $k$), the smoother the axial profile of the vacuum magnetic field (the smaller the parameter $q$), and the smaller the mirror ratio $K$. In the previous paper \cite{Kotelnikov+2022NF_62_096025} we did not aim to calculate $\Lambda_{\min}$ exactly, but simply chose the minimum values of $\Lambda$ from the available list of discrete values for which we calculated $\beta_ {\text{crit}}$.


In the present work, we tried to calculate $\Lambda_{\min}$ more accurately. The difficulty of such calculations is that the coefficients of the LoDestro equation are singular at the point $z=0$ in the limit $\beta \to 1$. On the other hand, we showed above that the LoDestro equation is inapplicable in this limit, since the paraxial approximation is violated at $\beta \to 1$. Nevertheless, we searched for the roots of the dispersion equation \eqref{5:17} with respect to $\Lambda $ for four sufficiently large beta values, $\beta=\{0.999, 0.9999,0.99999, 0.999999\}$. The values of $\Lambda _{\min}$ calculated in this way are given in tables \ref{tbl:minLambda} for $\beta =0.999$.


A cursory analysis of these tables shows that $\Lambda_{\min}$ noticeably increases as the parameters $q$ and $R$ increase, that is, both as the magnetic mirrors steepen and as the anisotropy decreases. On the contrary, $\Lambda_{\min}$ decreases with an increase in the parameter $k$, that is, with a steepening of the radial pressure profile, and this decrease is especially noticeable at lower anisotropy. The same tendency towards a decrease in $\Lambda_{\min}$ is observed as the mirror ratio $K$ decreases.

The smallest value $\Lambda _{\min}=1.0000000000024$ was found in the case $\beta =0.999999$, $k=\infty $, $q=2$, $K=16$.


\section{Search for the roots of a nonlinear equation}\label{A3}

The search for the roots of Eqs.~\eqref{4:10}, \eqref{5:17} and \eqref{6:04} was initially carried out using the \texttt{FindRoot} utility built into the Wolfram \emph{Mathematica}$^{\copyright}$ library. \texttt{FindRoot} searches for a root near an initial guess $\beta_{\text{start}}$ passed to it. Success or failure in finding the root with this utility depends very much on luck in choosing $\beta_{\text{start}}$. Therefore, as an additional means of searching for roots, the \texttt{RootSearch} package was included, which was developed by Ted Ersek \cite{Ersek_RootSearch}. This package contains a utility of the same name that searches for all roots within a given interval.

In some intermediate version of our numerical code, a root found by the \texttt{RootSearch} utility was passed as $\beta_{\text{start}}$ to the \texttt{FindRoot} utility to recheck the result of calculation of $\beta_{\text{crit}}$. In rare cases, when only one of the two utilities found a solution to Eqs.~\eqref{4:10}, \eqref{5:17} or \eqref{6:04}, the code was analyzed in order to improve it. In particular, the formulas that Wolfram \emph{Mathematica}$^{\copyright}$ obtained when calculating the integrals $a^{2}$ and $\mean{\overline{p}}$ were improved. For example, in the formula \eqref{A1:33f} for $a_{4}^{2}$, the number of Appell hypergeometric functions was reduced from four to one, and in the formula \eqref{3:43f} for $\mean{\overline{p}}_{4}$, from six to two. A side result of such code optimization was the ``expulsion'' of complex numbers in the intermediate calculations of the integral \eqref{4:10}. Due to rounding errors for such numbers, it sometimes happened that the integral \eqref{4:10} took on a complex value with a small but finite imaginary part. In such cases, \texttt{FindRoot} skipped the root of Eq.~\eqref{4:10}. In cases where both \texttt{FindRoot} and \texttt{RootSearch} did not find a solution to Eq.~\eqref{4:10}, it was considered that the solution did not exist.

Unfortunately, after all this effort, the \texttt{FindRoot} and \texttt{RootSearch} bundle of utilities kept missing roots of  Eqs.~\eqref{4:10}, \eqref{5:17}, \eqref{6:04}. The problem of losing roots was solved by applying the ideas formulated in the post of user with the nickname matheorem in the \texttt{mathematica} forum on the \texttt{stackexchange.com} portal. Using his code sample \cite{matheorem2020} and some of the code from Ted Yersek's \texttt{RootSearch} \cite{Ersek_RootSearch} package, we wrote a \texttt{XRS} package that replaced the \texttt{FindRoot} and \texttt{RootSearch} utilities.



%

\end{document}